\documentclass[journal]{IEEEtran}
\usepackage{ifpdf}
\usepackage{xcolor}

\usepackage{cite}

\usepackage{amssymb}

\usepackage[pdftex]{graphicx}
 \usepackage[font=footnotesize]{caption}
  \usepackage[font=footnotesize]{subcaption}
\ifCLASSINFOpdf
  \usepackage{graphicx}
  \usepackage[font=footnotesize]{caption}
  \usepackage[font=footnotesize]{subcaption}
\else
\fi
\usepackage{amsmath}
\usepackage{array}
\newcolumntype{P}[1]{>{\centering\arraybackslash}p{#1}}
\usepackage{url}

\begin{document}
%
\title{Low-Power (1T1N) Skyrmionic Synapses for Spiking Neuromorphic Systems}
%
%
%

\author{Tinish Bhattacharya, Sai Li, Yangqi Huang, Wang Kang, Weisheng Zhao and Manan Suri
\thanks{Manuscript received 30 March 2018. This research was supported in part by the Department of Science and Technology, Science and Engineering Research Board - Extramural Research Grant, Government of India Grant and Faculty Interdisciplinary Research Project - IIT-D grant and in part by the National Natural Science Foundation of China (Grant 61627813 and 61571023), and the International Collaboration Project (Grant 2015DFE12880 and B16001). T. Bhattacharya and S. Li contributed equally to this work. (Corresponding Author: Manan Suri and Weisheng Zhao).}
\thanks{T. Bhattacharya and M. Suri are with the Department of Electrical Engineering, Indian Institute of Technology - Delhi, 110016 India (e-mail: manansuri@ee.iitd.ac.in).}
\thanks{S. Li is with Shenyuan Honors College and BDBC, Fert Beijing Research Institute of Beihang University, Beijing, 100191 China.}
\thanks{Y. Huang, W. Kang and W. Zhao are with BDBC, Fert Beijing Research Institute, Beihang University, Beijing, 100191 China (e-mail: weisheng.zhao@buaa.edu.cn).}}

\maketitle

\begin{abstract}
In this work, we propose a `1-transistor 1-nanotrack' (1T1N) synapse based on  movement of magnetic skyrmions using spin polarised current pulses. The proposed synaptic bit-cell has 4 terminals and fully decoupled spike transmission- and programming- paths. With careful tuning of programming parameters we ensure multi-level non-volatile conductance evolution in the proposed skyrmionic synapse. Through micromagnetic simulations, we studied in detail the impact of programming conditions (current density, pulse width) on synaptic performance parameters such as number of conductance levels and energy per transition. The programming parameters chosen used all further analysis gave rise to a synapse with 7 distinct conductance states and  1.2 fJ per conductance state transition event. Exploiting bidirectional conductance modulation the 1T1N synapse is able to undergo long-term potentiation (LTP) \& depression (LTD) according to a simplified variant of biological spike timing dependent plasticity (STDP) rule. We present subthreshold CMOS spike generator circuit which when coupled with well known subthreshold integrator circuit, produces custom pre and post-neuronal spike shapes, responsible for implementing unsupervised learning with the proposed 1T1N synaptic bit-cell and consuming $\sim$ 0.25 pJ/event. A spiking neural network (SNN) incorporating the characteristics of the 1T1N synapse was simulated for two seperate applications: pattern extraction from noisy video streams and MNIST classification. 

\end{abstract}


%
\IEEEpeerreviewmaketitle

\section{Introduction}
%
%
%
%
\IEEEPARstart{H}{ardware} implementation of biologically inspired neuromorphic systems has gained a lot of interest in the past few years \cite{IBM_PCM,ICM_Truenorth}. Neuromorphic engineering aims to achieve intelligent computing by taking inspiration from complex neural/synaptic circuits and their biophysical mechanisms, using nano-electronic/magnetic devices \cite{b1,b2}. The nanodevices when made to follow certain conductance (weight) modulation rules within a connected network lead to equivalent trained neural networks that may be used for different training and inference tasks \cite{b4,b8}. Various emerging non-volatile nanodevices are currently being investigated for synaptic applications; Conductive-Bridge Memory (CBRAM) \cite{b4}, Phase Change Memory (PCM) \cite{b5}, Magnetic-Tunnel Junction (MTJ) \cite{b6}, domain walls \cite{sengupta2016hybrid} and oxide based resistive switching memory (OxRAM) \cite{b7,b8}. Among emerging spintronic nanodevices, magnetic skyrmions have gained a significant amount of interest owing to their small size, topological stability and ultralow current densities \cite{b9,b25}. Skyrmions are topologically stable spin textures which have been experimentally observed in Heavy metal (HM)-Ferromagnetic Metal (FM) heterostructure and bulk ferromagnets \cite{b24,b23}. They have been proposed as possible post-Moore candidates for logic \cite{b18}, \cite{b17} and storage \cite{b19} applications. They have also been proposed for emulating the properties of leaky-integrate and fire neurons in \cite{li2017magnetic,chen2018compact}. However for a device to behave as analog (multilevel) synapse, it should have multiple programmable non-volatile conductance states \cite{b7}. Controlled conductance variation using skyrmions on nanotracks was first reported in \cite{b26}. However in \cite{b26} the conductance states reported were a strong function of the inter programming-pulse delay, thus making the intermediate synaptic weights transient or not truly non-volatile. Also to the best of our knowledge no work has yet shown a pathway for implementing online unsupervised learning with multilevel skyrmion on nanotrack synapses.

In this work we present a modified nanotrack structure which makes the overall synapse a 4-terminal device and completely decouples the spike transmission and programming paths. This simplifies the synaptic circuit and results in a 1T1N like configuration. We then study in detail the impact of two different programming parameters: pulse width and pulse amplitude, on the number of distinct conductance levels and energy per transition of the proposed Skyrmion on Nanotrack (Sk-N) synapse. Using this optimized nanotrack and programming parameters we propose a ’1 Transistor/1 nanotrack’ (1T1N) Skyrmion on Nanotrack (Sk-N) based synaptic bit-cell that changes the conductance of the synapse according to a modified version of the biological unsupervised STDP learning rule \cite{ambrogio2016neuromorphic}. We also design peripheral subthreshold CMOS neuron circuit comprising of following blocks: current scaling circuit, Differential Pair Integrator adopted from \cite{qiao2015reconfigurable,livi2009current}, comparator and a custom spike generator unit. The current scaling circuit used for this work is based on the Gilbert normalizer circuit \cite{liu2002analog} and was recently shown to work with two-terminal differential memristive synapses \cite{nair2017differential}. In \cite{nair2017differential}, the circuit was included inside each synapse bit-cell. In this work, we show how the same circuit can be used effectively for 4-Terminal non-differential Sk-N based synapses by coupling it with each post-neuron rather than the synaptic bit-cells, thus saving silicon footprint. The spike generator unit is able to generate custom pre and post-neuronal spikes that lead to the modified STDP learning rule. On combining all the circuits we demonstrate spike transmission, neuronal integration and simplified STDP based conductance modulation in a single synaptic bit-cell through circuit simulations. The circuit simulations were done in CADENCE Spectre, using TSMC 65 nm technology node PDK. The characteristics of our 1T1N synaptic bit-cell and programming circuit were used in system-level simulations to demonstrate online unsupervised learning in a Spiking Neural Network (SNN) in the MATLAB based neuromorphic hardware simulator, MASTISK \cite{mastisk}. The task involved pattern recognition in noisy video streams and multi-class classification of handwritten digits (MNIST).

\section{Skyrmion on Nanotrack}

\subsection{Skyrmion Physics}

The creation of a skyrmion can be attributed to the competition between ferromagnetic exchange coupling and Dzyaloshinskii–Moriya interaction (DMI) in magnetic systems. These systems have breaking or lacking inversion symmetry in bulk lattices \cite{b23} or at the interface of thin films \cite{b24} respectively. The DMI interaction between two neighbouring atomic spins in a lattice having large Spin - Orbit Coupling (SOC) is give by:
\begin{align}
\label{eq:DMI}
\boldsymbol{H} = &-\boldsymbol{D} . (\boldsymbol{S_{i}}\times\boldsymbol{S_{j}}),
\end{align}

where D denotes the DMI vector, and S$_{i}$ and S$_{j}$ are the
spins on site i and j , respectively. In this work we are mainly concerned with skyrmions created in interface of ultrathin films (also known as nanotracks) comprising of a Ferromagnetic-Metal (FM) layer deposited on a Heavy-Metal (HM) layer. In such systems spin polarised current is used to disturb the magnetization equilibrium of the metal layer and induce topological transition of the magnetic textures. This when facilitated by DMI leads to creation of skyrmions. Skyrmions on nanotracks can be moved by either Current-In-Plane (CIP) or Current-Perpendicular-to-Plane (CPP). In this work the latter case is implemented due to its energy efficiency \cite{sampaio2013nucleation}. According to this method a charge current flowing through the HM layer induces a vertical spin current due to Spin-Hall Effect (SHE) which is responsible for moving the skyrmions in the FM layer by exerting spin torque. Due to the non-linear spin texture of skyrmions, there will be definite change in the site-dependent spin mixing of magnetic states in the ferromagnetic environment. This can be detected using Tunneling Magneto Resistance (TMR) measurements \cite{gould2004tunneling}.

\subsection{Skyrmion on Nanotrack Synapse}

\begin{figure}[!t]
\centering
\includegraphics[width=\columnwidth]{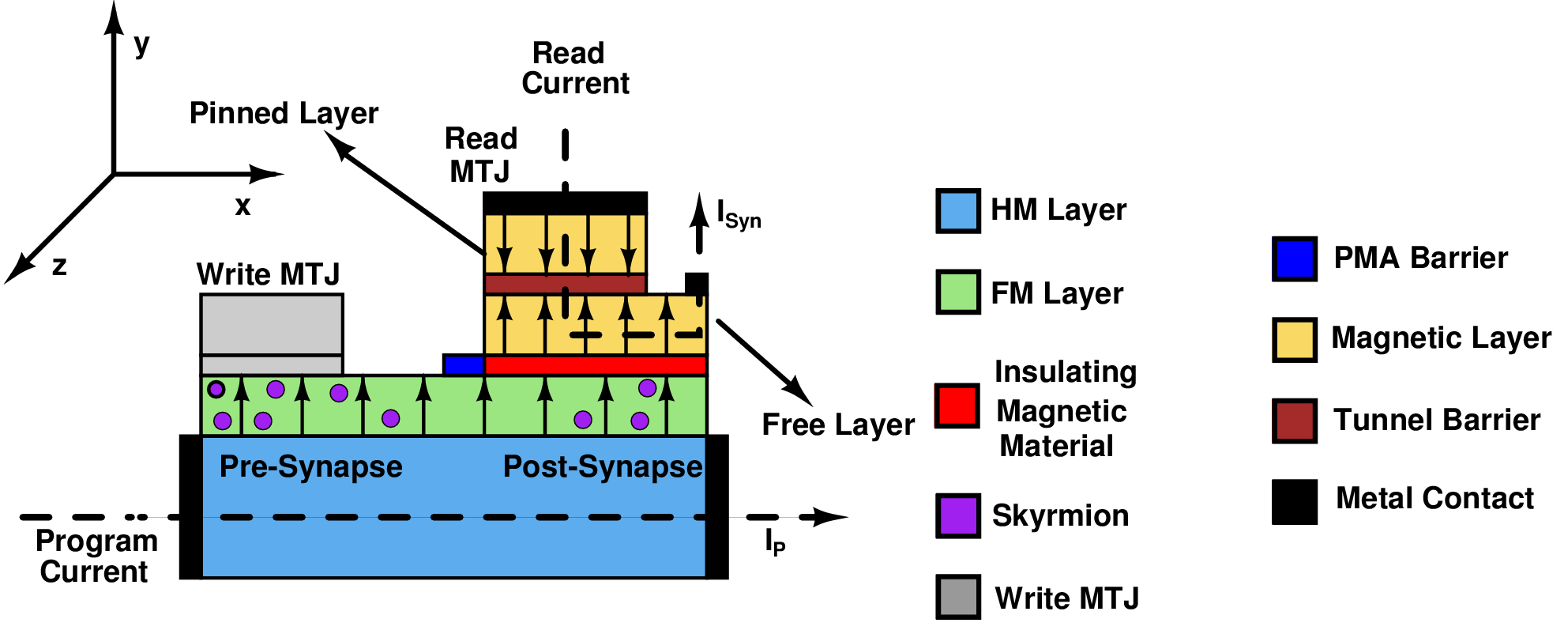}
\caption{Schematic of proposed skyrmion on nanotrack based synaptic device with completely decoupled read and program paths.}
\label{skyrmion1}
\end{figure}

The concept of skyrmion on nanotrack synapse used for this work is shown in Fig. \ref{skyrmion1}. It has ultrathin metallic films comprising of a Ferromagnetic Metal (FM) layer deposited on top of a Heavy Metal (HM) layer. The FM layer has Perpendicular Magnetic Anisotropy (PMA) in the direction specified by $\hat{y}$ (Fig. \ref{skyrmion1}). A metallic capping layer with a stronger PMA than the FM layer is used as an energy barrier for the skyrmions, thus seperating the nanotrack into two parts: pre-synapse and post-synapse. Please note that the terminologies pre-synapse and post-synapse have been adopted from \cite{b26}. They just refer to the two sides of the synapse separated by the PMA barrier and have no relation with pre-neuron and post-neuron. MTJ like structures are used at either ends of the nanotrack for nucleating (write-MTJ) and reading (read-MTJ) skyrmions respectively. The write-MTJ (pinned layer is the portion of FM layer below it) is used to introduce spin-polarised current into the pre-synapse region in order to create skyrmions. The read-MTJ has its free layer separated from the FM layer with an insulating magnetic material that allows magnetic coupling but prevents current flowing across it. The pinned and free layers of the Read-MTJ are separated by non-magnetic spacer. This decouples the read and write/program paths of the nanotrack completely.
The benefit obtained by doing so is discussed in Section \ref{2t1n}. The read current (I$_{Syn}$) flowing through the write-MTJ (shown in Fig. \ref{skyrmion1}) will be proportional to the conductance that is determined by the spin configurations of the post-synaptic FM layer and the free and pinned layers of the read-MTJ. Since the spin direction of the FM layer is opposite to that of the pinned layer of read-MTJ, it is in Anti-parallel (AP) state in the absence of skyrmions. As the net spin of a skyrmion in the $\hat{y}$ direction is zero, its presence in the post-synaptic region is equivalent to removal of a portion of the FM layer along with its spin. This leads to a less AP nature of the read-MTJ and therefore an increase in conductance. Therefore higher the number of skyrmions in the post-synapse, higher is the conductance of the post-synapse and the corresponding read current flowing through it. Programming current I$_{P}$ (charge current) when passed through the HM layer, the spin current caused by it as a consequence of Spin Hall Effect (SHE) is responsible for moving the skyrmions from one end of the nanotrack to the other. 

\subsection{Micro-Magnetic Simulation Framework}

The skyrmion-based artificial synaptic device was numerically simulated by 3D magnetization dynamics in the Object Oriented MicroMagnetic Framework (OOMMF) software \cite{b22} which used the extended Landau-Lifshitz-Gilbert equation including spin transfer torques as follows \cite{wiesendanger2016nanoscale}:

\begin{align}
\label{eq:LLGS-CPP}
\frac{d\boldsymbol{m}}{dt} = &-\left|\gamma_{0}\right|\boldsymbol{m}\times\boldsymbol{h}_{\text{eff}}+\alpha(\boldsymbol{m}\times\frac{d\boldsymbol{m}}{dt}) \\ \notag
&+\frac{u}{t}\boldsymbol{m}\times (\boldsymbol{p}\times \boldsymbol{m}),
\end{align}

where $\gamma$ is the gyromagnetic ratio, $\boldsymbol{m}$ =$\frac{\boldsymbol{M}}{\boldsymbol{M_{S}}}$ is the reduced magnetization, $\boldsymbol{h}_{eff}$ = $\frac{\boldsymbol{H}_{eff}}{\boldsymbol{M_{S}}}$ is the reduced effective field, $\alpha$ is the Gilbert Damping Factor, $u=|\frac{\gamma_{0}\hbar}{\mu_{0}e}|\frac{jP}{2M_{\text{S}}}$, $\hbar$ is the reduced Planck constant, $j$ is the applied current density, $P=0.6$ is the spin polarization, $\mu_{0}$ is the permeability of free space, $e$ is the electron charge, $\boldsymbol{p}$ is the unit spin polarization direction and $\boldsymbol{p}=-\hat{y}$ was set for driving the skyrmions. The parameters were extracted from experiments \cite{wiesendanger2016nanoscale} for a 0.4 nm thick Co layer on a Pt substrate.

In order to calculate the conductance of the post-synaptic region Tunneling Magneto Resistance (TMR) has been employed. This is done by dividing the read - MTJ into multiple 2 nm $\times$ 2 nm cells in the x-y plane and calculating the conductance of each cell by Julliere's model \cite{julliere1975tunneling}: 

\begin{align}
\label{eq:julliere1}
G = G_{0}(\frac{1+p^{2}\cos\theta}{1+p^{2}}) \\ \notag
   putting \ \theta \Rightarrow \pi - \theta \\ 
\label{eq:julliere2}  = G_{0}(\frac{1-p^{2}\cos\theta}{1+p^{2}}),
\end{align}

where G$_{0}$ is the conductance when the magnetization is perfectly
parallel to the reference layer, p is the spin polarization and $\theta$ is the magnetization of each cell with respect to the reference layer. However since in this work the spins of the reference layer of MTJ and FM layer are oppositely oriented, $\theta$ is replaced by $\pi - \theta$ leading to the form shown in Eq. \ref{eq:julliere2}. As a result the introduction of skyrmions in the post-synapse leads to $\theta$ being less than $\pi$ which gives a higher value of G according to Eq. \ref{eq:julliere2}. 

\begin{table}[htbp]
  \centering
  \caption{Device parameters for the HM-FM heterostructure}
    \begin{tabular}{|c|c|p{8.215em}|}
    \hline
    Parameter & Description & \multicolumn{1}{c|}{Value} \\
    \hline
    M$_{s}$    & Saturation Magnetization & \multicolumn{1}{c|}{580 kA/m} \\
    \hline
    A     & Exchange Constant & \multicolumn{1}{c|}{15 pJ/m} \\
    \hline
    D     & DMI Factor & \multicolumn{1}{c|}{3 mJ/m2} \\
    \hline
    $\alpha$ & Gilbert Damping Factor & \multicolumn{1}{c|}{0.3} \\
    \hline
    K     & Magnetic anisotropy & \multicolumn{1}{c|}{0.8 MJ/m$^{3}$} \\
    \hline
    P     & Spin Polarization & \multicolumn{1}{c|}{0.6} \\
    \hline
    t$_{f}$    & Thin Film Thickness & \multicolumn{1}{c|}{1 nm} \\
    \hline
    l $\times$ w & Nanotrack Length and Width & 820 nm $\times$ 280 nm \\
    \hline
    $\rho$  & Resistivity of HM layer & 100 $\mu$$\Omega$cm \cite{b27} \\
    \hline
    \end{tabular}%
  \label{tab:device_params}%
\end{table}%

\section{Nanotrack Programming parameters}
\label{opt}
In the nanotrack simulations it was found that when skyrmions try to flow across the PMA barrier, there exists a competition between; driving force due to current, repulsive force due to nanotrack edge, PMA barrier and inter-skyrmion (Sk-Sk) interactions. If repulsive and driving forces are not optimized it leads to leaky movement of skyrmions across the barrier after the programming pulses are removed. This makes the conductance levels dependent on the inter-pulse delay as was the case in \cite{b26} and thus cannot be used for stable temporal long-term (LTP/LTD) non-volatile synaptic emulation. Each skyrmion crossing over to the post-synapse region increases the conductance of the read path. Thus, maximum number of conductance levels that can be achieved is the total number of skyrmions nucleated in pre-synapse. If multiple skyrmions enter the post-synapse under the influence of a single pulse then the effective number of distinct conductance states is reduced. The maximum number of stable skyrmions that are created during nucleation depends on the width of nanotrack \cite{b26}. Therefore the number of conductance levels is proportional to the width of the nanotrack. Increasing the length, allows more room for sequential movement of skyrmions in response to driving currents. However a large width and length would also incur synaptic area overheads. Therefore dimension of the nanotrack was fixed at: 820 nm $\times$ 280 nm $\times$ 1 nm. It allowed nucleation of 17 skyrmions in the pre-synapse during initiation. The spin polarisation value was set to 0.6 so as to improve the control of driving current on the movement of skyrmions across the PMA barrier. The micromagnetic simulation parameters are listed in Table \ref{tab:device_params}. 

\begin{figure}[!t]
\centering
\begin{subfigure}{0.85\linewidth}
  \centering
  \includegraphics[width=1\linewidth]{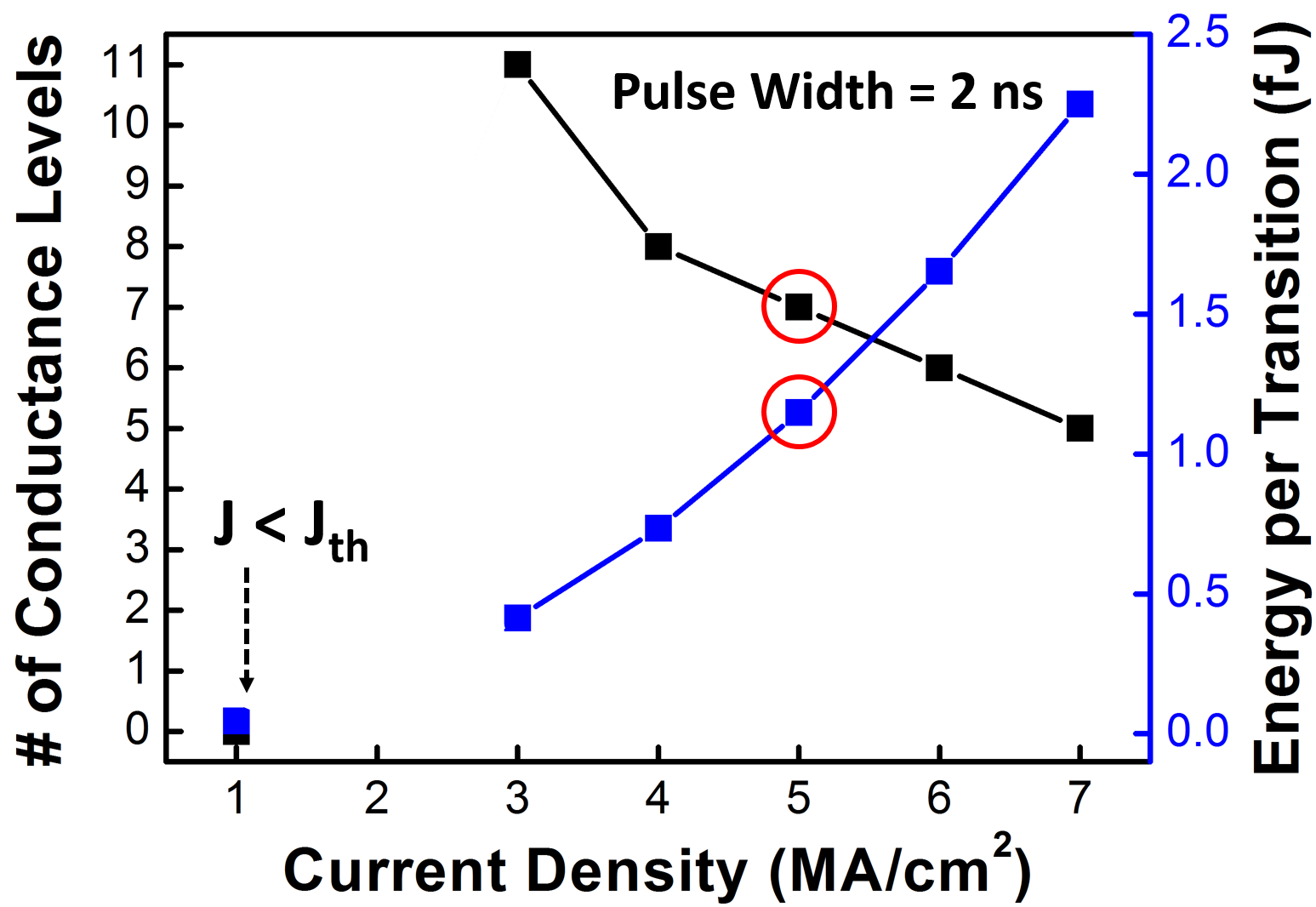}
  \caption{}
  \label{fig:sub1}
\end{subfigure}
\begin{subfigure}{0.85\linewidth}
  \centering
  \includegraphics[width=1\linewidth]{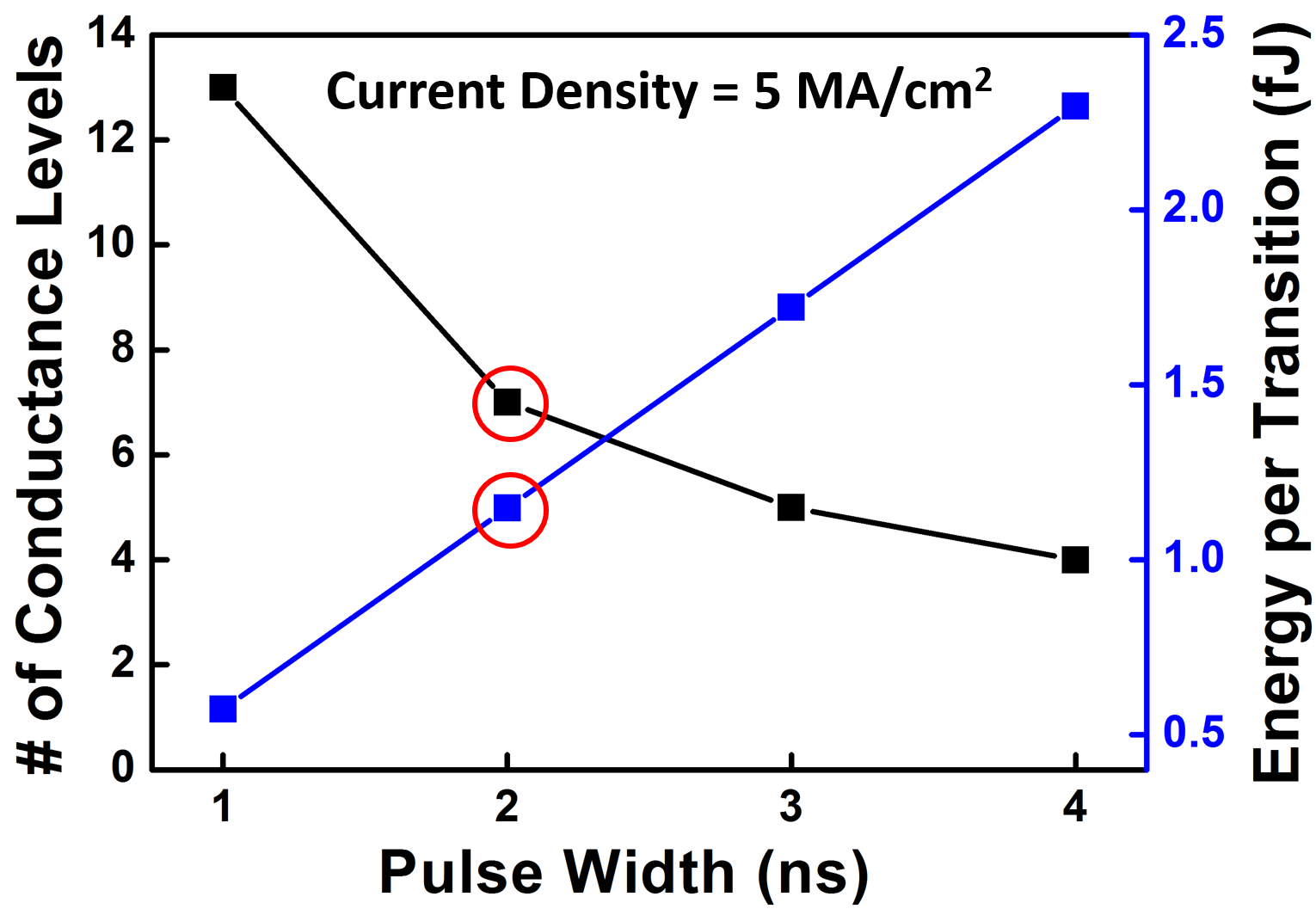}
  \caption{}
  \label{fig:sub2}
\end{subfigure}
\caption{Variation of number of conductance states and energy per conductance state transition with (a) current density and (b) pulse width of programming pulse. The red circle represents the programming parameter value chosen for other analysis in this work.}
\label{fig:master_opt}
\end{figure}

\begin{figure}[!b]
\centering
\includegraphics[scale = 0.40]{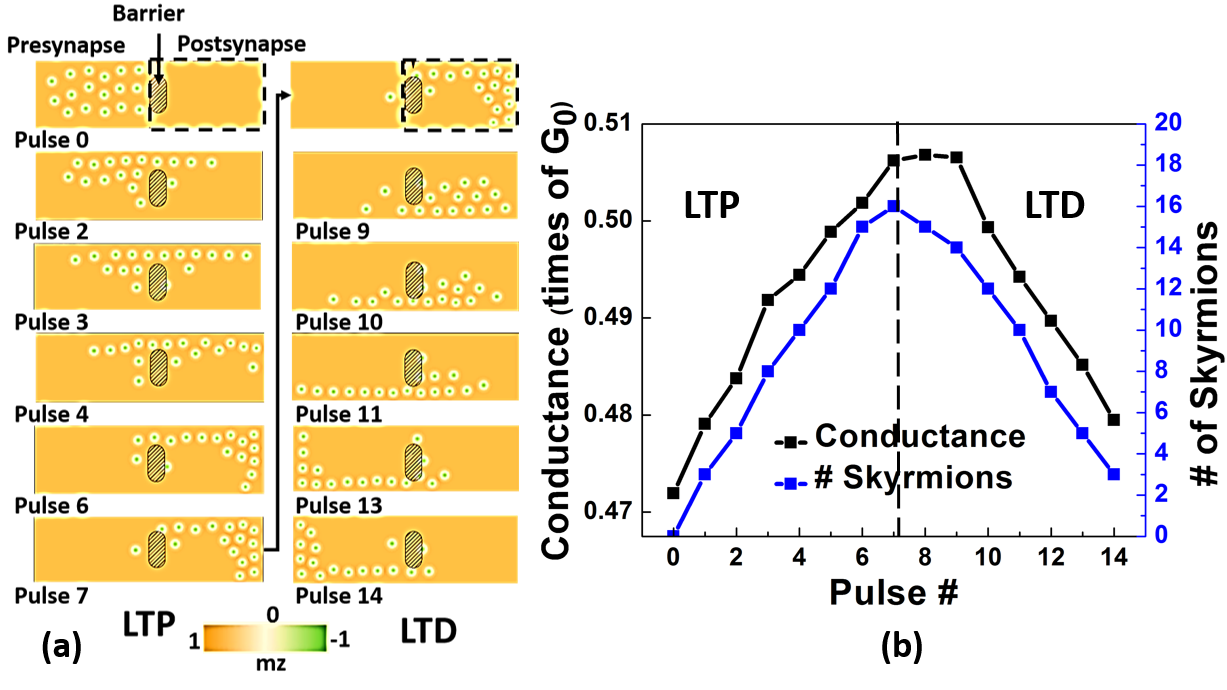}
\caption{(a) Micromagnetic simulations showing non-volatile conductance modulation of our skyrmionic synapse. First 7 pulses (5 MA/cm$^{2}$, 2 ns width) move the skyrmions from pre to post-synapse region (LTP) and the next 7 vice-versa (LTD). (b) Evolution of post-synapse conductance.}
\label{conductance}
\end{figure}

The variation of conductance levels and programming energy per pulse with current density is shown in Fig. \ref{fig:master_opt} (a). As the current density increases, the number of conductance states decrease. This is because with higher values of current, the events when multiple skyrmions cross the barrier increases. However if the current density is lower than a certain threshold (J$_{th}$) then it fails to move the skyrmions across the barrier. The current pulse width was kept at a constant value of 2 ns. The energy values are calculated using Equation \ref{eq:energy}:
\begin{align}
\label{eq:energy}
E_{Spike} = \rho\times t\times l\times w\times J^{2}\times T_{W},
\end{align}
where $\rho$ is the resistivity of Pt thin film \cite{b27}, t and w are the thickness and width of HM layer and J and T$_{W}$ are the current density amplitude and width of the pulses used for programming. The variation of conductance levels and energy for different programming pulse widths is shown in Fig. \ref{fig:master_opt} (b). Current density of the pulses used was 5 MA/cm$^{2}$. As the pulse width was decreased, the time window allowed for multiple skyrmions to cross the barrier decreased and therefore the number of distinct conductance levels increased. It was also found that just after a pulse ended, there was some transient component in the skyrmions' velocity. The maximum inter pulse delay required for the skyrmion movement across the barrier to stabilise was found to be 5 ns. This puts a maximum limit on the operating frequency of neuromorphic systems built with this nanotrack. In all the simulations only those conductance states are reported which were non-volatile and did not involve any skyrmion crossing barrier after removal of programming pulse.
The programming parameters used in this work are: current density = 5 MA/cm$^{2}$ and pulse width = 2 ns and has been circled in red in Fig. \ref{fig:master_opt}. Using these parameters, consecutive potentiating (depressing) pulses which moved skyrmions from pre to post-synaptic (post to pre-synaptic) region were applied. The position of skyrmions in the nanotrack after different pulses are shown in Fig. \ref{conductance} (a) whereas the conductance and post-synaptic skyrmion population variation with pulses is shown in Fig. \ref{conductance} (b). 
The conductance modulation curve is nearly linear and symmetrical in both LTP and LTD phases, which is a desirable property in electronic synapses \cite{sung2018effect}.

\begin{figure*}
\centering
\begin{subfigure}{.5\textwidth}
  \centering
  \includegraphics[scale=0.1]{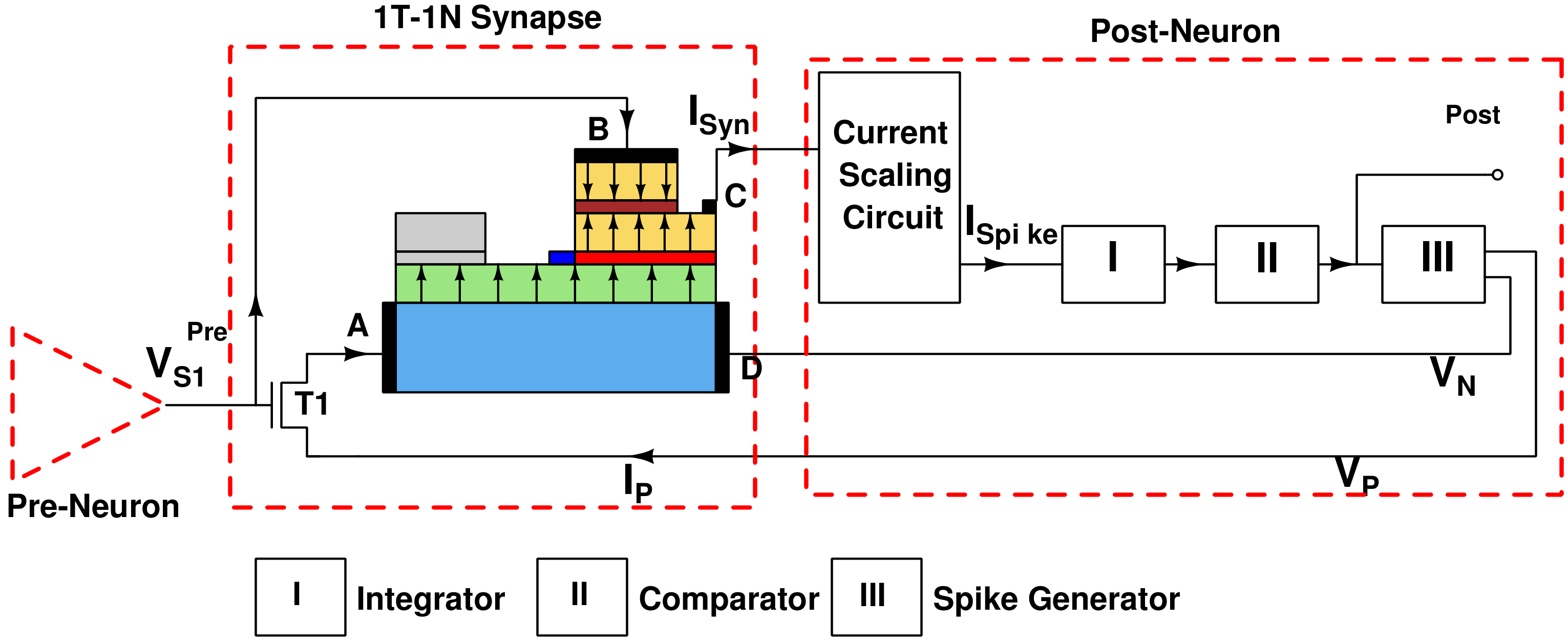}
  \caption{}
  \label{fig:sub1}
\end{subfigure}%
\begin{subfigure}{.5\textwidth}
  \centering
  \hbox{\hspace{2cm} \includegraphics[height=4.4cm,keepaspectratio]{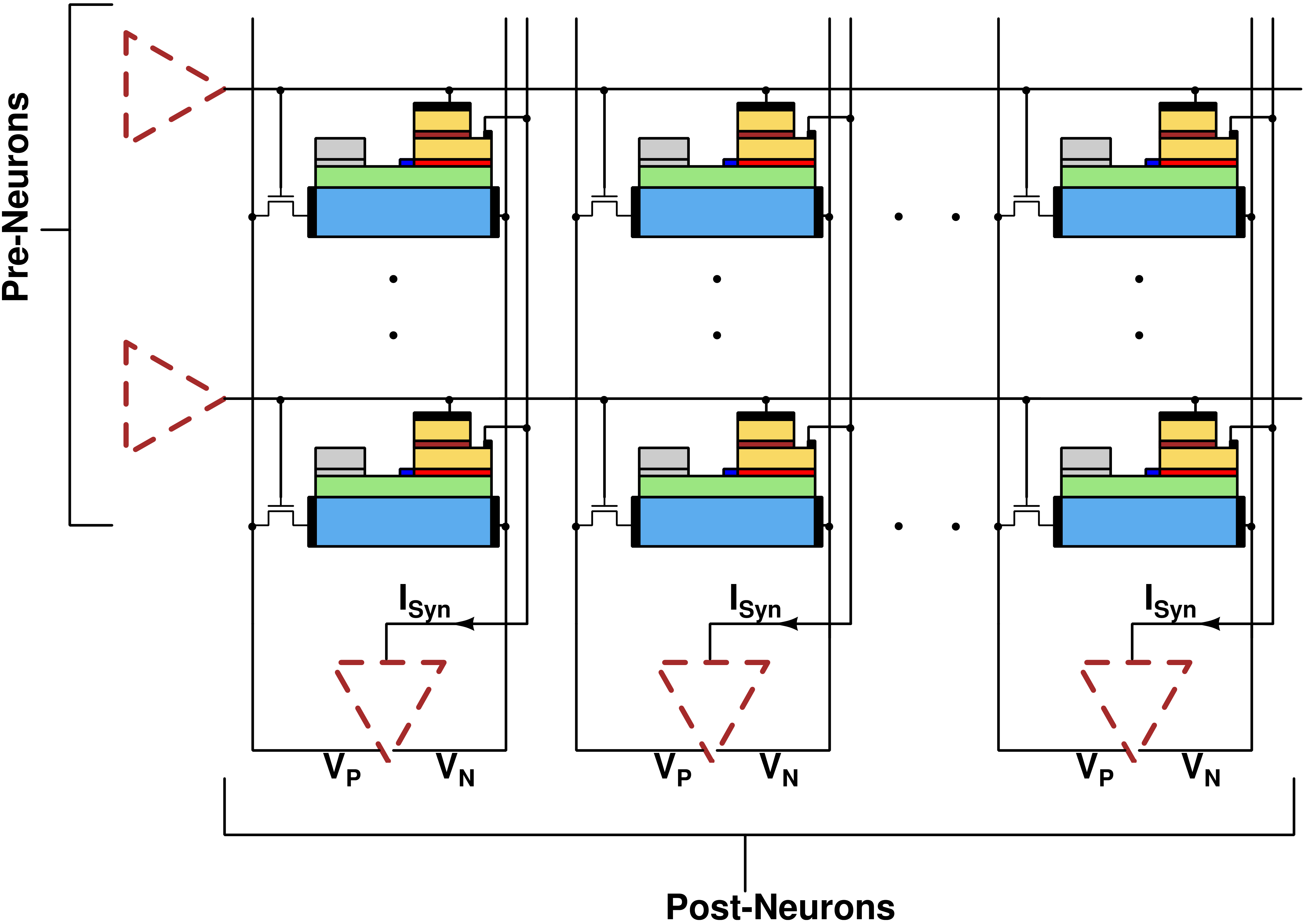}}
  \caption{}
  \label{fig:sub2}
\end{subfigure}
\caption{(a) Proposed 1T1N skyrmionic synapse and different components of post-neuron, (b) 1T1N synaptic bit-cells in crossbar arrangement.}
\label{fig:master_synapse}
\end{figure*}

\section{1T1N Synapse and Simplified STDP}
\label{2t1n}
\begin{figure}[!t]
\centering
\includegraphics[width=0.8\columnwidth]{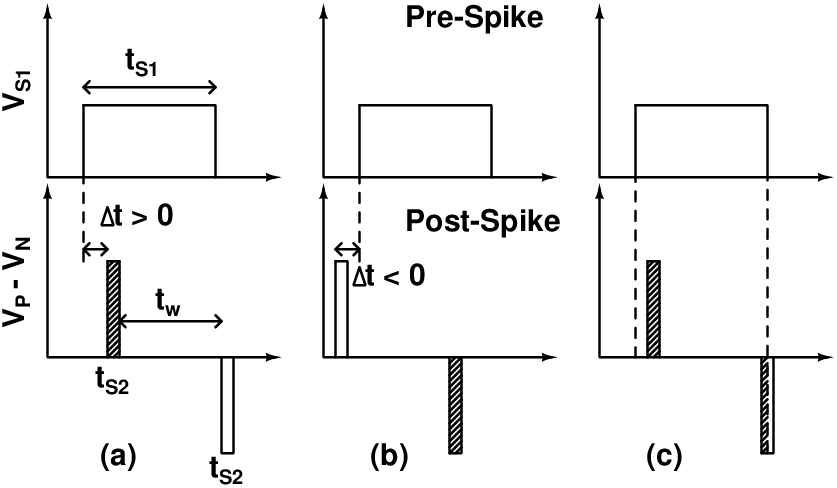}
\caption{Proposed pre and post-spike pulse shapes and their different orientation in time leading to (a) LTP, (b) LTD and (c) combination of LTP and LTD.}
\label{spikes}
\end{figure}

\begin{figure}[!t]
\centering
\includegraphics[scale = 0.37	]{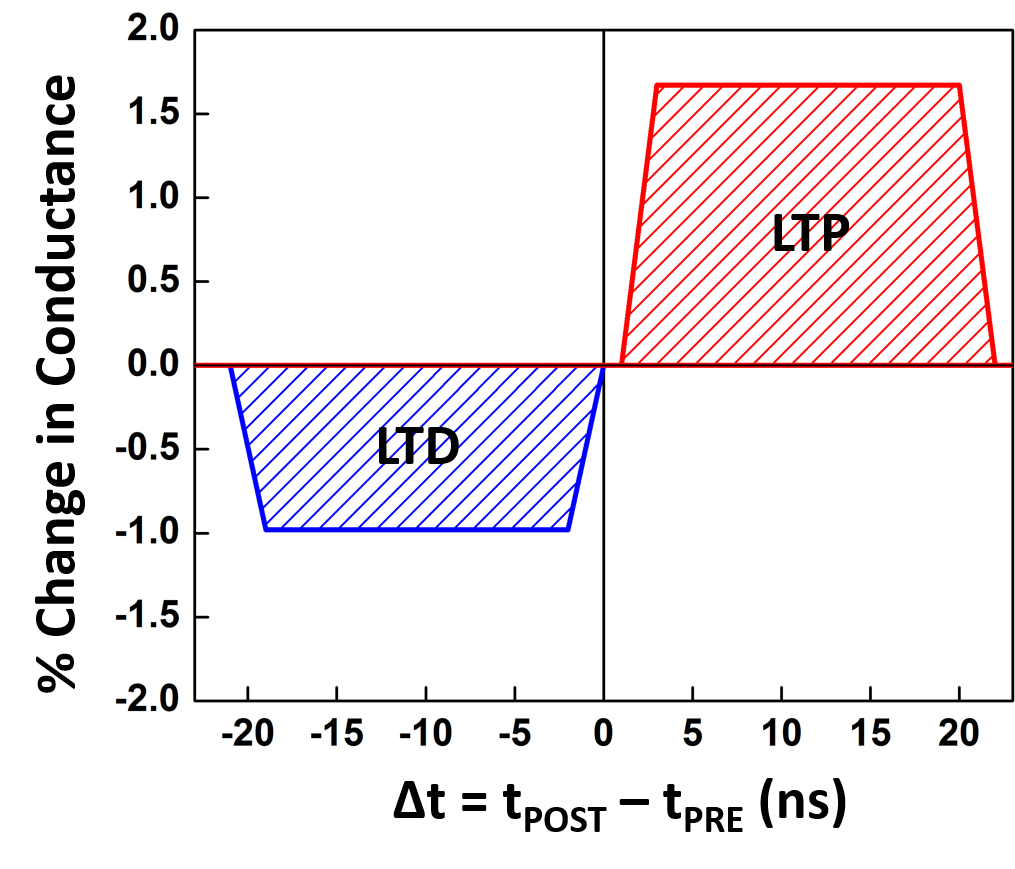}
\caption{Simplified STDP characteristics used in our proposed design. The percentage change in conductance is shown with respect to an initial state where 5 skyrmions are present in post-synaptic region. }
\label{stdp}
\end{figure}

The proposed 1T1N synaptic circuit along with the pre and post-neurons are shown in Fig. \ref{fig:master_synapse} (a). The proposed pre/post-neuronal spikes (programming methodology) are shown in Fig. \ref{spikes}. In our proposed synaptic design there are four possible modes of operation; (1) idle, (2) spike transmission, (3) potentiation and (4) depression. The different components of a neuron used in this work are: Current scaling circuit, integrator, comparator and a spike generator. The circuit and functioning of each of these components are discussed in detail in Section \ref{ref1}. Spike transmission occurs when there is a pre-spike i.e. V$^{Pre}_{S1}$ is HIGH. The pre-spike voltage applied across the terminals B and C result in a current proportional to the conductance of the synapse (I$_{Syn}$), flowing out of C and entering the post neuron through the current scaling circuit. Consequently a current I$_{Spike}$ enters the integrator and keeps getting integrated in the form of voltage (V$_{mem}$) across a capacitor. When V$_{mem}$ crosses a threshold, two different spikes are generated: V$_{P}$ - V$_{N}$ which is the post-neuronal spike for the synapse under consideration and V$^{Post}_{S1}$ which is the pre-neuronal spike for the next layer of synapses. Although during spike transmission the gate of T1 receives a HIGH voltage, if there is no post-neuronal spike it will be in cutoff and no significant current will flow between A and D. In the case when only post spike occurs, the synaptic circuit remains in idle mode. This is because there is no current flow through the synapse as the transistor T1 is off. Plasticity (LTP or LTD) occurs when both the pre and post-spikes have temporal coincidence. When the positive part of post-spike coincides with the pre-spike as shown in Fig. \ref{spikes} (a), LTP occurs (conductance of read-MTJ increases). At the onset of the pre-spike, T1 is in cutoff and spike transmission takes place from B to C. As soon as the post-spike arrives, T1 starts conducting and a current (whose magnitude depends on HM layer resistance and V$_{P}$ - V$_{N}$) flows from A to D, moving the skyrmions to the right. Its worth noting that spike transmission takes place between B and C even when the post-neuronal spike is driving skyrmions across the barrier in the HM layer. This is the unique advantage provided by our 4 Terminal decoupled read-program nanotrack. In case of a 3-terminal nanotrack \cite{sengupta2016hybrid,b26} separate transistors would be required for switching off spike transmission during programming mode, since both operations take place through a common node. LTD occurs when the negative part of the post-spike coincides with the pre-spike (Fig. \ref{spikes} (b)). This results in current flowing from D to A, thus moving the skyrmions from  post to pre-synaptic region and thus reducing the conductance of read-MTJ. For certain range of temporal spacing between pre and post-neuronal spikes, portions of both the positive and negative parts of the post-neuronal spike coincides with the pre-neuronal spike (shown in Fig. \ref{spikes} (c)). In such a case both LTP and LTD would take place in degrees proportional to their extent of their overlap. Based on the micromagnetic simulations and nanotrack parameters described in the previous section, the characteristics of pre/post neuron spikes used are; t$_{S2}$ = 2 ns, t$_{S1}$ = 22 ns and t$_{w}$ = 17 ns. Therefore the temporal conditions for which LTP, LTD and both occur are: 3 $\leqslant$ $\Delta$t \textless 22, -21 \textless $\Delta$t $\leqslant$ -2 and -2 \textless $\Delta$t \textless 3 respectively, where $\Delta$t = t$_{post}$-t$_{pre}$. The resultant STDP characteristic curve showing the percentage change in conductance for different values of $\Delta$t is shown in Fig. \ref{stdp}. The chosen spike scheme leads to synaptic programming energy consumption of $\sim$ 1.2 fJ/ per spike, (Equation \ref{eq:energy}). The arrangement of our proposed 1T1N synaptic bit-cell in crossbar arrangement is shown in Fig. \ref{fig:master_synapse} (b). All bit-cells in a column share the same post-neuron whereas all those in a row share the same pre-neuron. Therefore currents from all the synaptic bit-cells in a single column that receive pre-neuronal spike contribute to I$_{Syn}$ that enters the post-neuron below. All the bit-cells in a column receive post-neuronal spike across V$_{P}$ and V$_{N}$, but only those are programmed whose transistor is ON due to simultaneous presence of pre-spike. Also note that due to the decoupled write and read paths in our proposed synaptic structure, conductance modulation and spike transmission can both occur simultaneously, thus leading to faster unsupervised learning. Whereas in the structure proposed in \cite{b26} due to a shared node between the two paths only one function (spike transmission or conductance modulation) could occur at a time. Considering that the same learning rule is being implemented, an extra transistor would be required to switch off the spike transmission path when the programming pulses are being applied in that case. 




\begin{figure}[!t]
\centering
\includegraphics[width=\columnwidth]{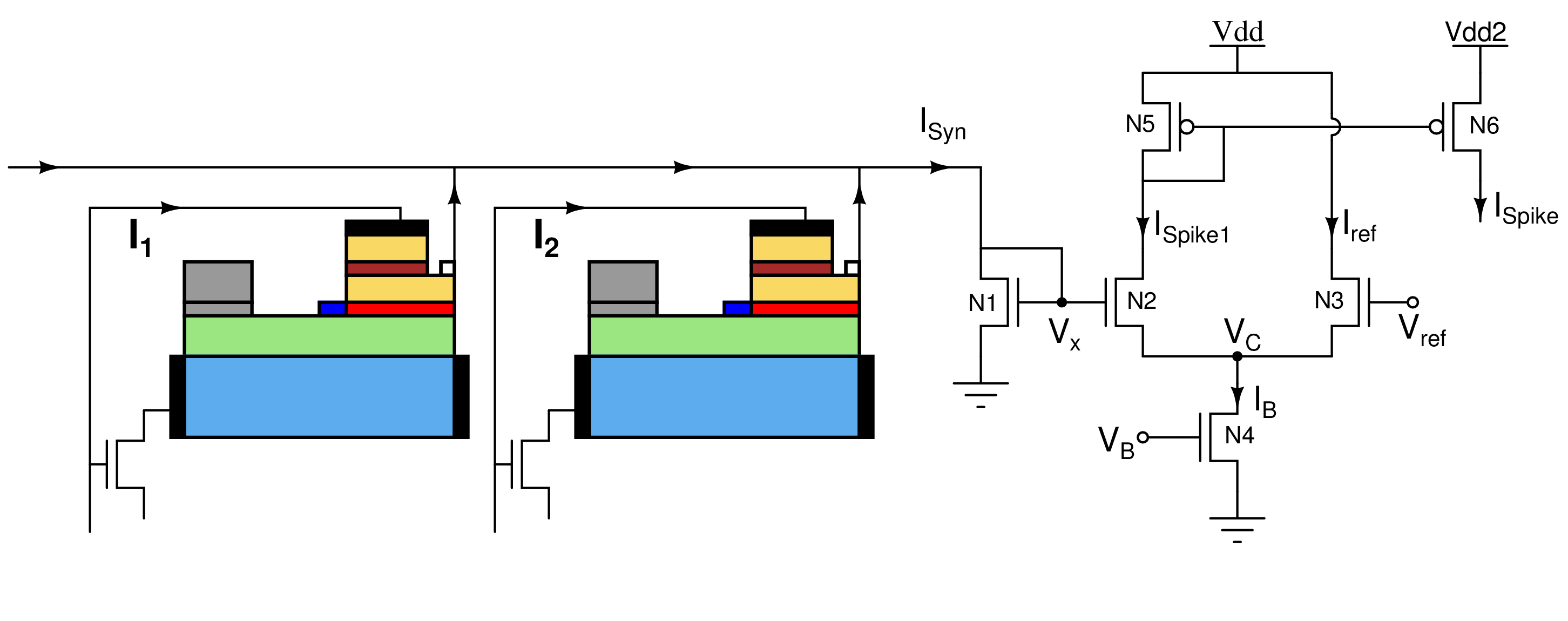}
\caption{Subthreshold current scaling and normalizing block used in our neuron circuit. The sum of current from synapses in previous layer (I$_{1}$, I$_{2}$ .. etc) enter the circuit block as I$_{Syn}$ and the resultant output current I$_{Spike}$ is passed to the integrator stage of the neuron.}
\label{dpi}
\end{figure}

\section{Subthreshold CMOS neuron circuit}
\label{ref1}
In this section we present the schematics, functioning and outputs of different building blocks of the CMOS based neuron circuit. All simulations were done in CADENCE Spectre, using TSMC 65 nm technology node PDK. The various circuit parameters used for simulation has been detailed in Table \ref{tab:circuit_params}. 

\subsection{Current Scaling Circuit}
\label{ref11}
The CMOS circuit involved in scaling and normalizing the current entering the post-neuron during spike transmission is shown in Fig. \ref{dpi}. The total current from synapses I$_{Syn}$, in the previous layer enter the post-neuron via this current scaling circuit block as I$_{Spike}$. The transistors N2-N6 operate in their subthreshold regime. This is ensured by making the bias current (I$_{B}$) controlled by V$_{B}$ and is very small, of the order of a few hundreds of nano-amperes. In that case the currents I$_{Spike1}$, I$_{ref}$ and I$_{B}$ are given by equation \ref{eq:dpi1}. 
The currents I$_{Syn}$ and I$_{Spike}$ are determined by equation \ref{eq:dpi1}:
\begin{align}
\label{eq:dpi1}
I_{Spike1} = I_{0}e^{\frac{kV_{x}-V_{C}}{V_{T}}}, \hspace{2mm} I_{ref} = I_{0}e^{\frac{kV_{ref}-V_{C}}{V_{T}}}, \hspace{2mm} I_{b} = I_{0}e^{\frac{kV_{B}}{V_{T}}}\\ \notag
\end{align}
On applying Kirchoff's law at node C, one can eliminate V$_{C}$ to get the relation between I$_{Spike1}$ and node voltage V$_{x}$, as shown in equation \ref{eq:dpi2}.
\begin{align}
\label{eq:dpi2}
I_{Spike1} = I_{b}\frac{e^{\frac{kV_{x}}{V_{T}}}}{e^{\frac{kV_{x}}{V_{T}}}+e^{\frac{kV_{ref}}{V_{T}}}},\\ \notag
\end{align}
Voltage V$_{x}$ is a function of the current I$_{Syn}$ and the exact relation between them depends on whether the transistor N1 operates in subthreshold or superthreshold regime. The PMOS transistors N5 and N6 are used as to make a current mirror that converts the sinking current flowing through N2 into a sourcing current flowing through N6 and therefore $I_{Spike1} \approx I_{Spike}$. As I$_{Spike}$ gets integrated in the neuron's integrator block, it's input impedance keeps increasing. Therefore in order to prevent the transistor N6 from going into cutoff region while trying to source the same amount of current into a larger impedance load, it's supply voltage (Vdd2) was kept at a slightly higher level (650 mV) as compared to Vdd1 (600 mV).

\begin{figure}[!h]
\centering
\includegraphics[scale = 0.3]{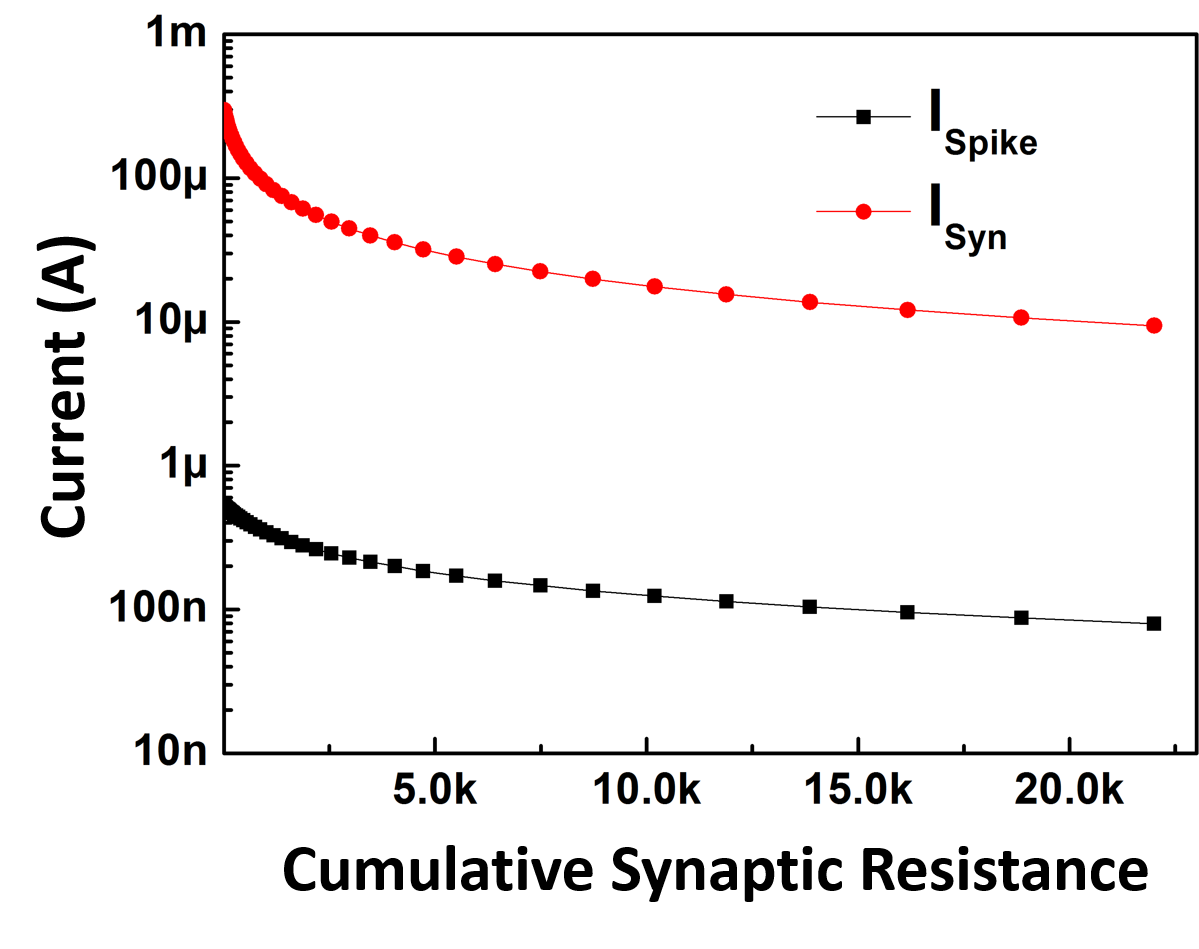}
\caption{Current response of the scaling and normalizing block for different values of cumulative resistance of the synapses connecting it to neurons in previous layer. }
\label{dpi_res}
\end{figure}

Variation of I$_{Syn}$ and I$_{Spike}$ with resistance of synapse is shown in Fig. \ref{dpi_res}. As more number of synapses (attached to different pre-neurons) are connected in parallel, the effective resistance of the path terminating in N1 decreases. Thus Fig. \ref{dpi_res} shows how values of I$_{Syn}$ and I$_{Spike}$ are affected when the number of synapses connecting a post-neuron to the previous neuron layer increases. The maximum resistance for which a current value has been plotted is 22 k$\Omega$, since that is the maximum resistance of a single nanotrack synapse when there are zero skyrmions in post-synaptic region.  

\begin{figure*}
\centering
\begin{subfigure}{.5\textwidth}
  \centering
  \includegraphics[width=1.2\linewidth]{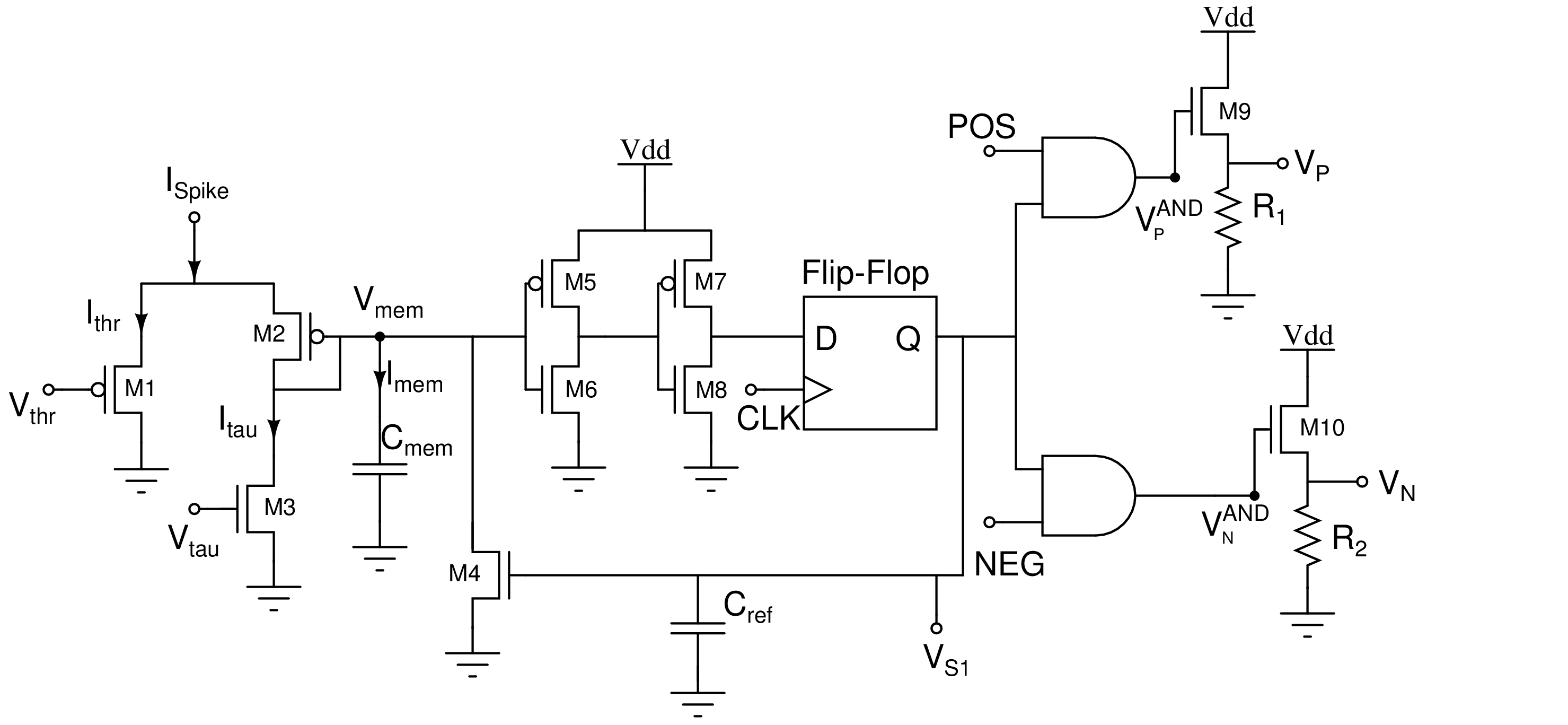}
  \caption{}
  \label{fig:sub1}
\end{subfigure}%
\begin{subfigure}{.5\textwidth}
  \centering
  \includegraphics[width=0.8\linewidth]{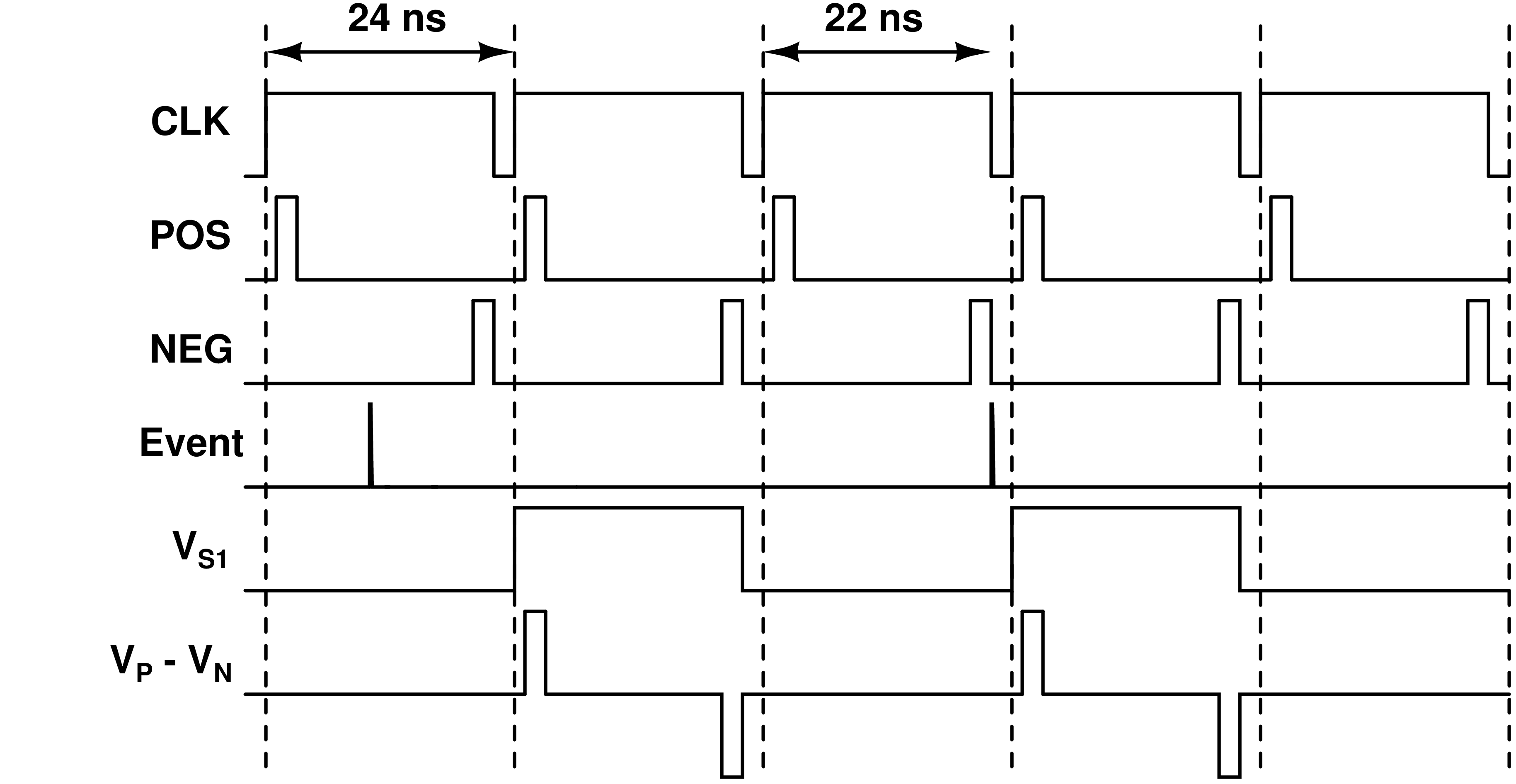}
  \caption{}
  \label{fig:sub2}
\end{subfigure}
\caption{(a) Subthreshold CMOS circuit of neuron used for this work, (b) Timing diagram showing the global signals (CLK, POS, NEG) and output spikes of the neuron (V$_{S1}$, V$_{P}$-V$_{N}$) for different temporal placements of spiking event.}
\label{fig:master_neuron}
\end{figure*}

\subsection{Integrator, Comparator and Spike Generator}
\label{ref12}
The CMOS circuit that constitutes the remaining of our post-neuron is shown in Fig. \ref{fig:master_neuron} (a). It starts with a Differential Pair Integrator filter for integrating incoming current signal (M1-M3). It forms the integrator block of our post-neuron. The voltage accumulated across the capacitance (C$_{mem}$) acts as the neuron's membrane potential. The inverters constituted by M5-M6 and M7-M8 play the role of comparator and spike event generating block. The circuit beyond it comprising of the Flip-Flop, AND gates, M9-M10 and R1-R2 is responsible for generating the spikes as specified in Fig. \ref{spikes}, whenever a spike event is encountered.

The expression governing the dynamics of a generic non-linear integrate and fire system is given by equation \ref{eq:generic_LIF} \cite{abbott1993asynchronous}. 
\begin{align}
\label{eq:generic_LIF}
\tau\frac{du}{dt} = F(u) + G(u)I\\ \notag
\end{align}
Assuming all transistors to be in subthreshold saturation and applying translinear principle, one gets the expression for I$_{out}$ as shown in equation \ref{eq:neuron_LIF1}, where the currents I$_{out}$, I$_{t}$, I$_{\tau}$ are given by equation \ref{eq:neuron_current} and $\tau = C_{mem}V_{T}/kI_{\tau}$.
\begin{align}
\label{eq:neuron_current}
I_{out} = I_{0}e^{\frac{kV_{mem}}{V_{T}}}, \hspace{2mm} I_{t} = I_{0}e^{\frac{kV_{thr}}{V_{T}}}, \hspace{2mm} I_{\tau} = I_{0}e^{\frac{kV_{tau}}{V_{T}}}\\ \notag
\end{align}
\vspace{-5mm}
\begin{align}
\label{eq:neuron_LIF1}
\tau\frac{dI_{out}}{dt} = \frac{I_{t}I_{out}I_{Spike}}{I_{\tau}(I_{out}+I_{t})} - I_{out}\\ \notag
\end{align}
On replacing $I_{out}$ with $I_{0}e^{\frac{kV_{mem}}{V_{T}}}$ (see equation \ref{eq:neuron_current}), one obtains the temporal dynamics of V$_{mem}$ shown in equation \ref{eq:neuron_LIF2}.
\begin{align}
\label{eq:neuron_LIF2}
\tau\frac{dV_{mem}}{dt} = \frac{I_{t}I_{Spike}}{I_{\tau}(I_{out}+I_{t})} - \frac{V_{T}}{k}\\ \notag
\end{align}
It is worth noting here that I$_{Spike}$ is the input current to the DPI integrator block and I$_{out}$ is a function of V$_{mem}$. Therefore both equation \ref{eq:neuron_LIF1} and \ref{eq:neuron_LIF2} represent non-linear integrate and fire dynamics (see equation \ref{eq:generic_LIF}) in variables I$_{out}$ and V$_{mem}$ respectively.

The threshold voltage of the neuron is determined by the switching voltage of the inverter (M5-M6). This can be controlled by changing the dimensions (W/L ratio) of M5 and M6. Therefore as V$_{mem}$ approaches the threshold, the output of the first inverter changes drastically to 0. In order to invert this output and provide spike events such that a high voltage corresponds to V$_{mem}$ reaching its threshold, a second inverter M7-M8 has been used. The second inverter also sharpens the response of the first inverter, thus resulting in positive spike events only when V$_{mem}$ crosses the threshold. 

In order to generate spikes of fixed pulse widths and desired shapes we make use of synchronous circuits, following the inverters. Three different clock signals have been used for this purpose: CLK, POS and NEG with widths as indicated in Fig. \ref{fig:master_neuron} (b). The Flip-Flop (FF) is used to generate the spike (V$_{S1}$) which acts as the pre-neuronal spike for the next layer of synapses. It's pulse width is same as that of CLK i.e. 22 ns. This spike is also connected to gate of M4 so as to discharge C$_{mem}$, thus reseting the accumulated membrane potential. C$_{ref}$ determines the time taken by C$_{mem}$ to discharge. Following this AND gates are used to make the outputs: V$^{AND}_{P}$ and V$^{AND}_{N}$ HIGH only when HIGH of POS and NEG overlap with HIGH of V$_{S1}$ respectively. The outputs from the AND gates are then connected to source follower circuits, so as to be able to drive larger loads. The final outputs of the circuit V$_{P}$ and V$_{N}$ are connected to the pre and post-synaptic regions of our nanotrack synapse. Therefore the differential voltage V$_{P}$ - V$_{N}$ represent the post-neuronal spike, used to drive skyrmions in the synapses in the previous layer. As we saw in Section \ref{opt}, different current amplitudes and pulse widths may lead to different conductance levels and energy dissipation in the synapse, one can control the amplitude of V$_{P}$ - V$_{N}$ by varying the values of R1, R2 and W/L ratios of M9 and M10, whereas the temporal parameters of the spikes (t$_{S1}$, t$_{w}$, t$_{S2}$ discussed in Section \ref{2t1n}) can be controlled by the pulse widths and duty ratios of CLK, POS, NEG. The time evolution of global clock signals, spike events and output spikes (V$_{S1}$ and V$_{P}$ - V$_{N}$) are shown in Fig. \ref{fig:master_neuron} (b). The temporal parameters of CLK, POS and NEG (shown in Fig. \ref{fig:master_neuron} (b)) were chosen so as to generate spikes with parameters as discussed in Section \ref{2t1n}. 

\begin{figure}[!t]
\centering
\includegraphics[width=\columnwidth]{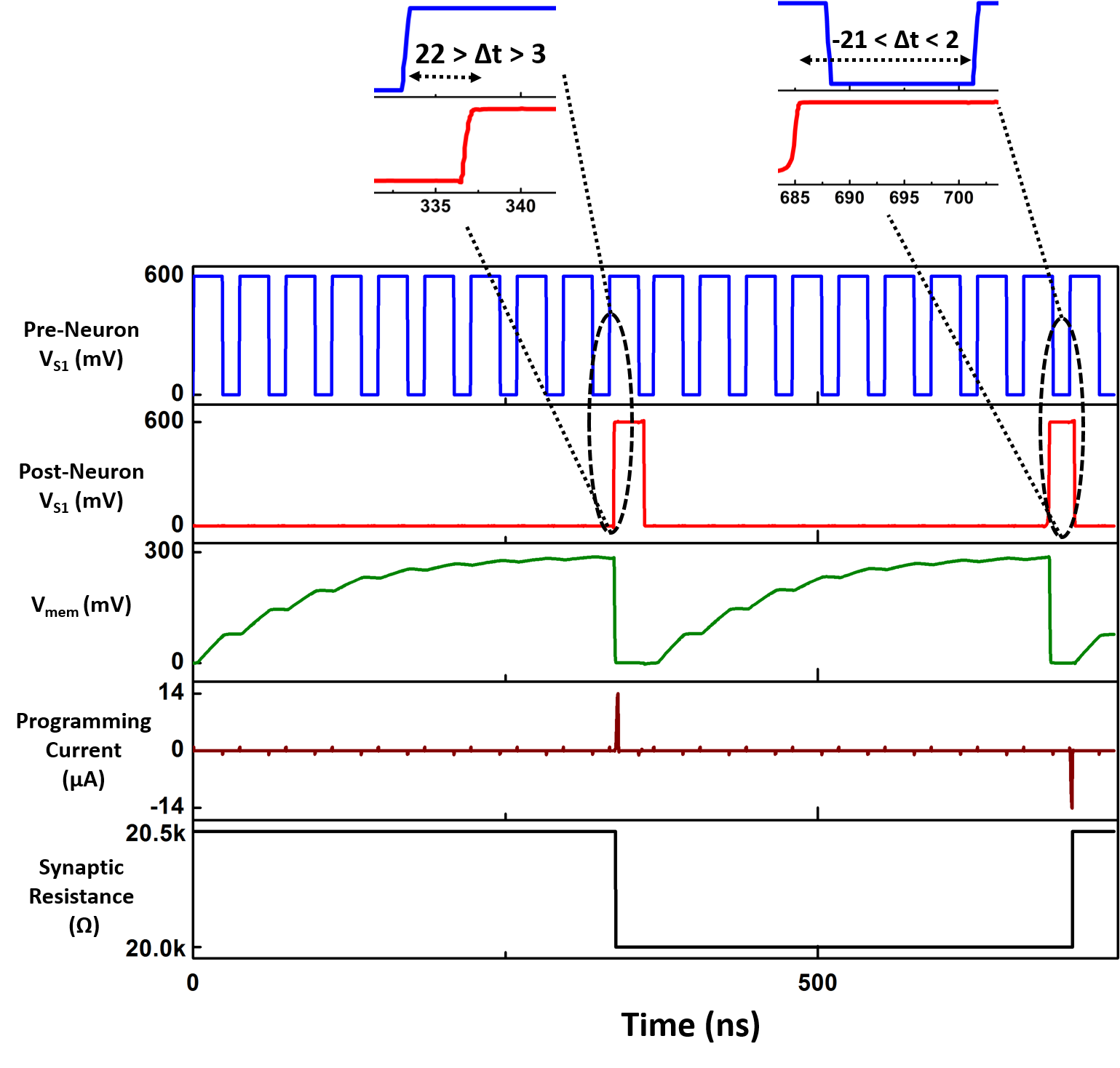}
\caption{Circuit simulation result showing pre-neuronal spikes, output spike and membrane potential of post-neuron, programming current flowing through synapse and its resistance when a single 1T1N synaptic bit-cell is connected between a pre-neuron and post-neuron.}
\label{neuron_res}
\end{figure}

In order to demonstrate the working of the proposed 1T-1N STDP powered skyrmionic synapse with neuron circuit, we considered a single synaptic bit-cell between a pre and post-neuron. The pre-neuron was modeled as a pulse source, generating (V$^{PRE}_{S1}$) with pulse width 23 ns and a frequency $\sim$ 27 MHz. The post-neuron was simulated with the circuit discussed in this section. The HM layer was modeled by a resistance of 3 k$\Omega$. The time evolution of V$^{POST}_{S1}$, V$_{mem}$ (membrane potential), programming current flowing through HM layer and resistance of read MTJ is shown in Fig. \ref{neuron_res}. It is worth noting here that the sign of the programming current shown in Fig. \ref{neuron_res} depends on the time ($\Delta$t) with which the pre and post-neuronal spikes are separated (exact dependence has been discussed in Section \ref{2t1n}). Accordingly current which is positive, negative or both might flow through the synapse, resulting in LTP, LTD or a combination of LTP and LTD respectively. The energy consumed in the neuron circuit (including the current scaling block, integrator, comparator and spike generation blocks) was found to be 0.25 pJ/spike for a supply voltage (V$_{dd}$) of 600 mV and neuron firing rate of $\sim$ 2.3 MHz.

\begin{table}[htbp]
  \centering
  \caption{Circuit parameters for CMOS Neuron Circuit}
    \begin{tabular}{|P{14.43em}|P{7.23em}|}
    \hline
    \textbf{Parameter Description} & \textbf{Value} \\
    \hline
     V$_{B}$   & 450 mV  \\
    \hline
     V$_{ref}$   & 450 mV \\
    \hline
     V$_{dd}$   & 600 mV \\
    \hline
     V$_{dd2}$   & 650 mV \\
    \hline
     V$_{thr}$   & 350 mV \\
    \hline
     V$_{tau}$   & 270 mV \\
    \hline
     C$_{mem}$   & 25 fF \\
    \hline
     C$_{ref}$   & 5 fF \\
    \hline
    W/L ratio of N1 & 10 $\mu$m / 65 nm \\
    \hline
    W/L ratio of N2-N6, M1-M3, M8 & 200 nm / 65 nm \\
    \hline
    W/L ratio of M5 & 240 nm / 65 nm \\
    \hline
    W/L ratio of M6 & 400 nm / 65 nm \\
    \hline
    W/L ratio of M7 & 400 nm / 65 nm \\
    \hline
     W/L ratio of M9-M10 & 16 $\mu$m / 65 nm \\
    \hline
     W/L ratio of T1 & 2.7 $\mu$m / 65 nm \\
    \hline
    \end{tabular}%
  \label{tab:circuit_params}%
\end{table}%

\section{Unsupervised learning with skyrmion synapse}
\begin{figure}[!t]
\centering
\includegraphics[scale = 0.2]{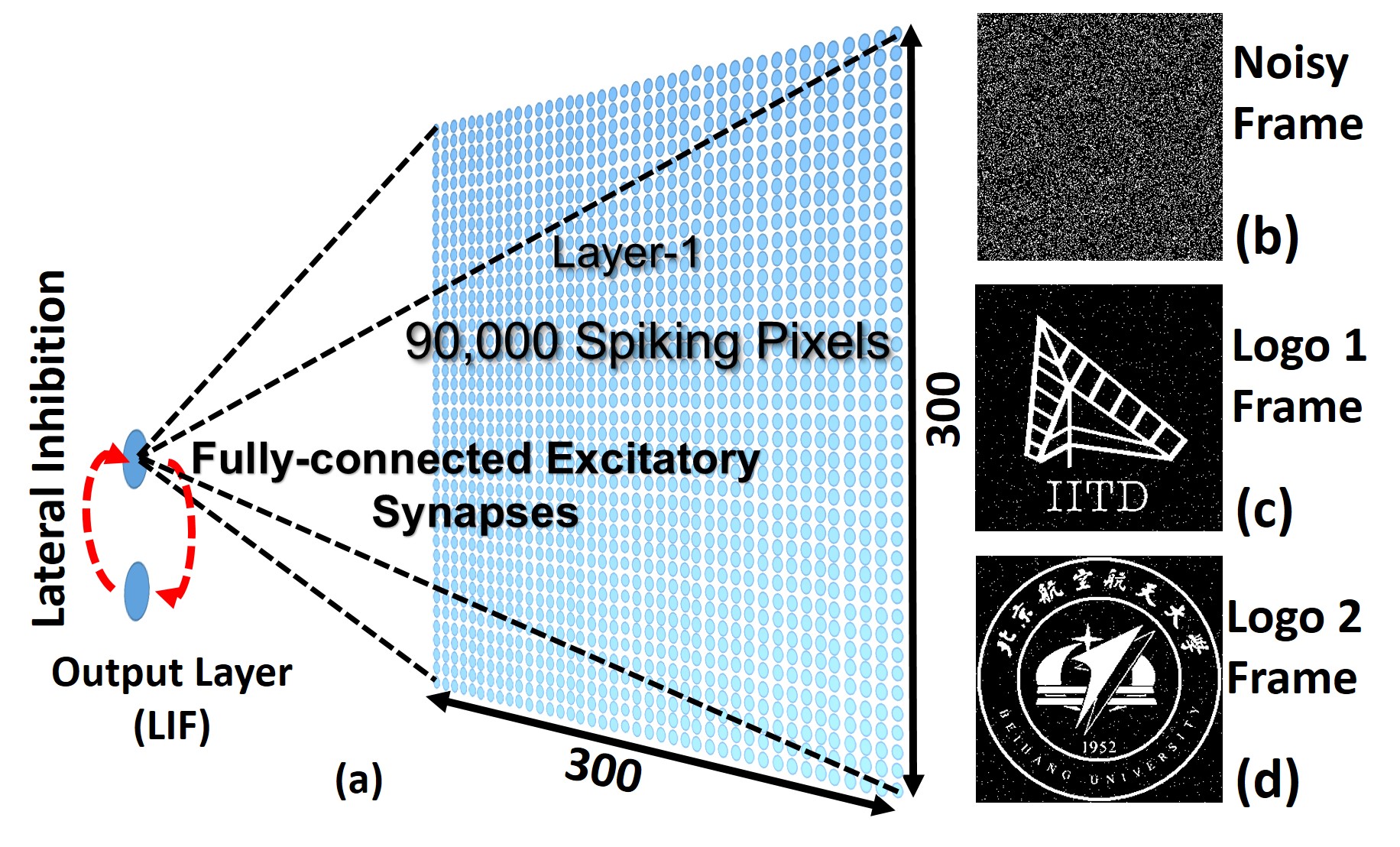}
\caption{(a) SNN topology simulated for this work (90 K neurons, 180 K synapses).(b-d) Frames of video dataset used for unsupervized learning application.}
\label{network}
\end{figure}

\begin{figure}[!t]
\centering
\includegraphics[scale = 0.3]{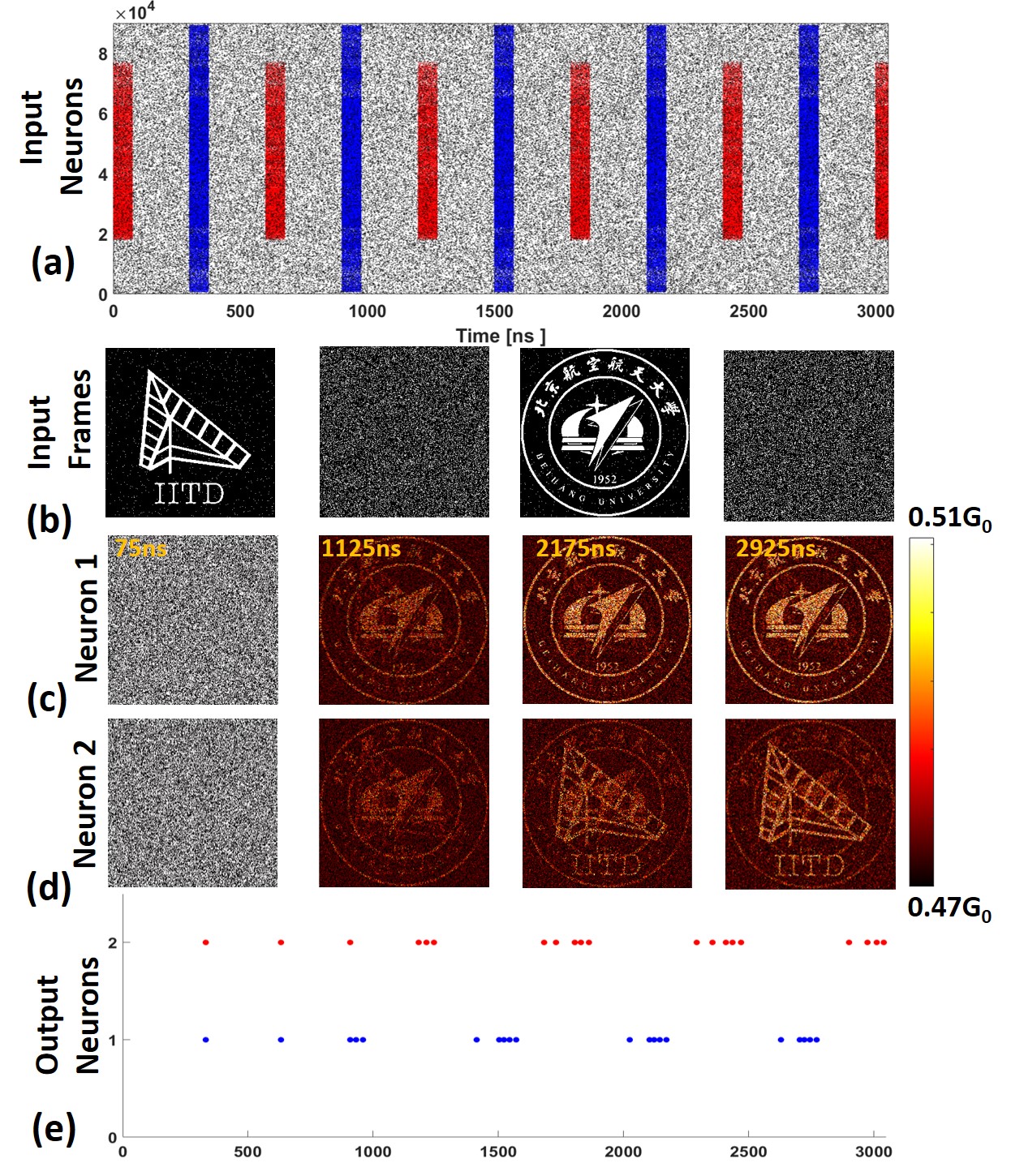}
\caption{(a) Input Rastor plot showing spiking activity of layer-1 neurons. (blue='IITD' logo, red = 'BUAA' logo occurence. (b) frames of input video data. (c)-(d) synaptic weight maps for Neuron 1 and 2 respectively, at timestamps: 75ns, 1125ns, 2175ns and 2925ns. (e) rastor plot for output neurons.} 
\label{fullresult}
\end{figure}

\begin{figure}[!h]
\centering
\includegraphics[scale = 0.35]{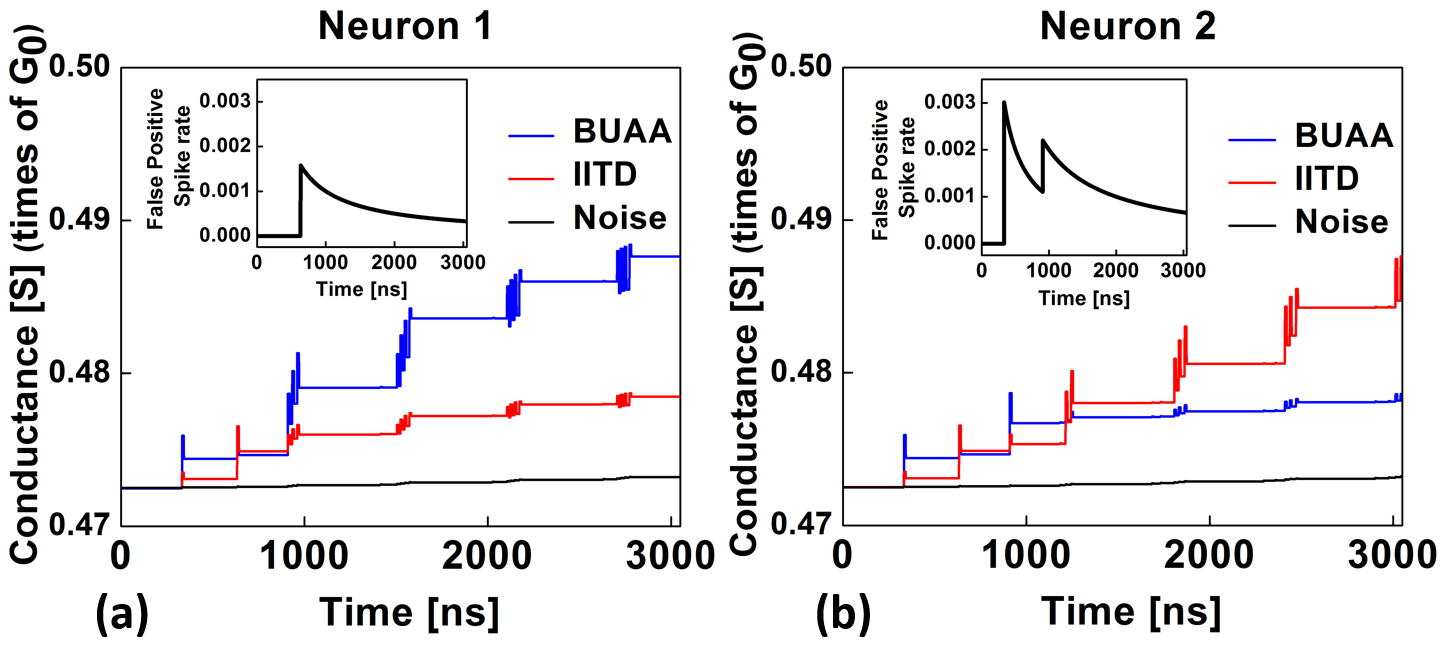}
\caption{Averaged conductance of the synapses corresponding to the input video for: (a) Neuron 1, and (b) Neuron 2. Inset shows false positive spike rate evolution with time.}
\label{cond2}
\end{figure}

In order to validate the working of our proposed skyrmion on nanotrack synaptic bit-cell for unsupervised learning we simulated two applications. In the first, we simulated a fully connected feed forward two-layer Spiking Neural Network (SNN) (see Fig. \ref{network}) powered by Spike-Timing Dependent Plasticity (STDP). All simulations were done on the MATLAB based neuromorphic simulator named MASTISK \cite{mastisk}. The network has 90,000 spiking pixels in the first layer and 2 LIF neurons in the output layer. The two layers are connected by excitatory synapses in an all-to-all fashion. The output neurons are connected to each other through inhibitory synapses. This is done to implement Winner-Take-All mechanism \cite{gupta2009hebbian}. Video stream comprising of binary 300 $\times$ 300 frames were used for training the SNN as shown in Fig. \ref{network} (b)-(d). The video stream comprises of mostly gaussian noise frames (see Fig. \ref{network} (b)) with two complex patterns (IIT-D logo and BUAA logo) embedded at different timestamps (Fig. \ref{network} (c)-(d)). The input layer encodes the frames into spikes using poisson spike encoding \cite{b14} with the mean firing rate proportional to the intensities of the pixels. Fig. \ref{fullresult} (a) shows the rastor plot of input neurons during the entire training period. Spiking activity of the neurons are denoted by dots in Fig. \ref{fullresult} (a) and the color denotes the kind of frame that was presented. Black dots represent noisy frame whereas the red and blue dots represent the frames containing the IIT-D logo and BUAA logo respectively. The input video frames at specific timestamps are illustrated in Fig. \ref{fullresult} (b). Fig. \ref{fullresult} (c) and (d) show the evolution of conductance of the synapses connecting the output neurons 1 and 2 to the input layer respectively, thus depicting the representations learned by each of the output neurons. Neuron 1 became selective to BUAA logo, while neuron 2 got selective to IIT-D logo. Initially the output neurons fire randomly, however from 1000 ns onwards as the occurrence of patterns increases the spiking becomes more specific to occurrence of a particular input pattern (Fig. \ref{fullresult} (e)). In order to study the degree of selectiveness of the neurons to different patterns, we separately noted the conductance of the synapses connecting the output neurons to different types of input pixels. The input pixels could either be pattern pixels and carry information for the two different patterns or be background pixel that do not carry information regarding any pattern. The averaged conductance of the synapses between the input pixels of the three types and two output neurons are shown in Fig. \ref{cond2}. The logo with maximum averaged conductance is the one for which a particular neuron gets selective. The difference between the conductances of the two patterns for a particular neuron depicts how well it can differentiate between the two patterns. The low conductance of the background synapses (noise) shows that the network is able to get selective to patterns and not background noise. The system achieves a low false positive spike rate of 6.5$\times$10$^{-4}$ (inset of Fig. \ref{cond2}). Since output neuron firing activity starts around 350 ns there is a transient in the false positive spike rate of either neuron. Ultra-low on-line unsupervised learning synaptic programming power consumption of $\sim$ 1 nW / synapse and neuron firing power consumption of $\sim$ 1.64 $\mu$W / neuron was achieved (Total programming events: $\sim$ 4.6$\times$10$^5$; Total post-neuron firing events: 40; synaptic programming energy: 1.2 fJ; CMOS neuron energy per firing event: 0.25 pJ; duration: 3050 ns; total synapses: 1.8$\times$10$^5$). The power takes in account both: energy spent in moving skyrmions in the nanotrack, and integration of I$_{Spike}$ in DPI integrator, comparator and the spike generation circuit over the entire duration of learning. 

\begin{table}[htbp]
  \centering
  \caption{MNIST Classification Accuracy for different synapses}
    \begin{tabular}{|P{14.43em}|P{7.23em}|}
    \hline
    \textbf{Synapse Type} & \textbf{Accuracy} \\
    \hline
   Redundancy = 2    & 73.8 \% \\
    \hline
   Redundancy = 4    & 82.17 \% \\
    \hline
    Redundancy = 6    & 84.97 \% \\
    \hline
    Ideal Synapse   & 85.5 \% \\
    \hline
    \end{tabular}
  \label{tab:mnist}
\end{table}

For the second application, we simulated a 3-layer SNN comprising of input, output (excitatory) and inhibitory neurons respectively inspired from \cite{diehl2015unsupervised}. The network was trained on 60,000 training images of the MNIST database. In the inference mode, the skyrmion on nanotrack characterized by its properties depicted in Fig. \ref{conductance}, was used as synapse. We varied the number of nanotracks connected in parallel to constitute a single synapse in order to see how performance varied with the redundancy. The different levels of redundancy and their corresponding classification accuracies have been shown in Table \ref{tab:mnist}. The synapse with redundancy = 6 almost reaches the performance level of ideal synapse (infinite dynamic range and conductance levels).

\begin{table*}[htbp]
  \centering
  \caption{Comparison with other synapses}
    \begin{tabular}{|P{6.5em}|P{6.855em}|P{6.43em}|P{6.645em}|P{6.355em}|P{6.355em}|p{6.07em}|p{6.57em}|}
    \hline
    \textbf{Device} & \textbf{Dimension} & \textbf{Programming Energy} & \textbf{Programming Time} & \textbf{Synapse Configuration} & \multicolumn{1}{p{6.355em}|}{\textbf{Terminals}} & \textbf{Type} & \textbf{Programming Neuron Spikes} \\
    \hline
    Ag-Si memristor \cite{jo2010nanoscale} & 100 nm X 100 nm & 25.2 pJ-403.2 pJ & 300 us & 1R    & 2     & Experimental & Identical \\
    \hline
    PCM \cite{kuzum2011nanoelectronic} & length: 500 nm, BE diameter: 75 nm & LTD (avg) = 13.5 pJ           LTP (avg) =  50 pJ & LTD transition: 75 ns LTP transition: 5 us  & 1R    & 2     & Experimental & Non-Identical \\
    \hline
    AlOx/HfO2 Bilayer RRAM \cite{woo2016improved} & 21 nm thick & LTP (avg) = 3.24 nJ   LTD (avg) = 4 nJ & 100 us & 1T1R  & 3     & Experimental & Identical \\
    \hline
    Domain Wall synapse \cite{sengupta2016hybrid} & 320 nm X 20 nm & 48 fJ/event & 1 ns   & 3T1R  & 5     & Simulation & Pres-Spikes = Complex exponential Post-Spikes =  identical pulses \\
    \hline
   1T1N Skyrmion Synapse [This Work] & 820 nm X 280 nm & 1.2 fJ/event & 2 ns   & 1T1R  & 4     & Simulation & Identical  \\
    \hline
    \end{tabular}%
  \label{tab:lit_comp}%
\end{table*}%

\section{Discussion}
Table \ref{tab:lit_comp} compares our proposed skyrmion synapse with other nanodevice based synapses. It clearly highlights the ultra-low energy (1.2 fJ/event) and sub-nanosecond time (2 ns) consumed in changing conductance states in our synapse. Since synapses based on nanotrack type structure (domain wall or skyrmion based) usually have 3 or more terminals, it becomes difficult to use nanotrack based synapses in 1R configuration (synaptic bit-cell containing only nanotrack and no transistor). For instance domain wall on nanotrack based synapse proposed in \cite{sengupta2016hybrid} had 3 transistors apart from the nanotrack in their synaptic bit-cell. However we show that using our proposed skyrmion on nanotrack configuration and neuron circuit one can implement a modified version of STDP learning rule with only 1 transistor along with the nanotrack in a bit-cell. 
A limitation of our synapse is the large dimensions of the nanotrack (1.5 \% larger than \cite{sengupta2016hybrid}). The length of the nanotracks determines the degree with which the skyrmions can freely move in the nanotrack without interacting with each other. It also has an impact on number of distinct conductance levels that can be achieved. In order to achieve same number of conductance states with shorter length it would require scaling down skyrmion sizes. In our simulations we used skyrmions with diameter $\sim$ 15 nm. Recently studies have shown that size of skyrmions can even be reduced to 1-3 nm \cite{wiesendanger2016nanoscale,romming2015field}. This opens possibilities of reducing the nanotrack length further without losing number of conductance levels. 

This work is based on simulations done holistically at the device, circuit and architectural level with parameters calibrated to experimental results \cite{wiesendanger2016nanoscale}. It clearly highlights the benefits of neuromorphic systems built with hybrid CMOS-skyrmion circuits.

\section{Conclusion}
We illustrate an approach for practical realization of multi-level synapse using hybrid skyrmion-CMOS based spiking neuromorphic systems based on simulations done at device, circuit and system level. Firstly through micromagnetic simulations we study in detail the impact of different programming parameters of a HM/FM nanotrack on various synaptic performance parameters and demonstrate true non-volatile multi-level conductance states, independent of inter-spike delay or frequency. We use a read-MTJ on top of the post-synaptic region of the nanotrack separated by insulating magnetic material from the FM layer beneath it. This makes our synapse 4-terminal with completely decoupled read and program paths. Leveraging the conductance modulation, we propose a 1T1N synaptic architecture and programming methodology to implement a modified version of the biological STDP rule, while consuming $\sim$ 1.2 fJ energy per programing event. We design a custom subthreshold synchronous spike generator circuit which when coupled with a current scaling circuit, Differential Pair Integrator and inverter based thresholding circuit, can perform the desired functionalities of a Leaky-Integrate and Fire neuron at extremely low energy (0.25 pJ/output spike). Unsupervised learning is demonstrated by simulating feed-forward SNN for pattern extraction and multi-class classification application. 
\ifCLASSOPTIONcaptionsoff
  \newpage
\fi



%
\bibliographystyle{IEEEtran}
\bibliography{IEEEabrv,main.bib}

\begin{thebibliography}{10}
\providecommand{\url}[1]{#1}
\csname url@samestyle\endcsname
\providecommand{\newblock}{\relax}
\providecommand{\bibinfo}[2]{#2}
\providecommand{\BIBentrySTDinterwordspacing}{\spaceskip=0pt\relax}
\providecommand{\BIBentryALTinterwordstretchfactor}{4}
\providecommand{\BIBentryALTinterwordspacing}{\spaceskip=\fontdimen2\font plus
\BIBentryALTinterwordstretchfactor\fontdimen3\font minus
  \fontdimen4\font\relax}
\providecommand{\BIBforeignlanguage}[2]{{%
\expandafter\ifx\csname l@#1\endcsname\relax
\typeout{** WARNING: IEEEtran.bst: No hyphenation pattern has been}%
\typeout{** loaded for the language `#1'. Using the pattern for}%
\typeout{** the default language instead.}%
\else
\language=\csname l@#1\endcsname
\fi
#2}}
\providecommand{\BIBdecl}{\relax}
\BIBdecl

\bibitem{IBM_PCM}
G.~W. Burr, R.~M. Shelby, S.~Sidler, C.~di~Nolfo, J.~Jang, I.~Boybat, R.~S.
  Shenoy, P.~Narayanan, K.~Virwani, E.~U. Giacometti, B.~N. Kurdi, and
  H.~Hwang, ``Experimental demonstration and tolerancing of a large-scale
  neural network (165 000 synapses) using phase-change memory as the synaptic
  weight element,'' \emph{IEEE Transactions on Electron Devices}, vol.~62,
  no.~11, pp. 3498--3507, Nov 2015, doi: 10.1109/TED.2015.2439635.

\bibitem{ICM_Truenorth}
P.~A. Merolla, J.~V. Arthur, R.~Alvarez-Icaza, A.~S. Cassidy, J.~Sawada,
  F.~Akopyan, B.~L. Jackson, N.~Imam, C.~Guo, Y.~Nakamura, B.~Brezzo, I.~Vo,
  S.~K. Esser, R.~Appuswamy, B.~Taba, A.~Amir, M.~D. Flickner, W.~P. Risk,
  R.~Manohar, and D.~S. Modha, ``A million spiking-neuron integrated circuit
  with a scalable communication network and interface,'' \emph{Science}, vol.
  345, no. 6197, pp. 668--673, Aug 2014, doi: 10.1126/science.1254642.

\bibitem{b1}
G.~Indiveri, E.~Chicca, and R.~Douglas, ``A vlsi array of low-power spiking
  neurons and bistable synapses with spike-timing dependent plasticity,''
  \emph{IEEE Transactions on Neural Networks}, vol.~17, no.~1, pp. 211--221,
  Jan 2006, doi: 10.1109/TNN.2005.860850.

\bibitem{b2}
C.~Mead, ``Neuromorphic electronic systems,'' \emph{Proceedings of the IEEE},
  vol.~78, no.~10, pp. 1629--1636, Oct 1990, doi: 10.1109/5.58356.

\bibitem{b4}
M.~Suri, O.~Bichler, D.~Querlioz, G.~Palma, E.~Vianello, D.~Vuillaume,
  C.~Gamrat, and B.~DeSalvo, ``Cbram devices as binary synapses for low-power
  stochastic neuromorphic systems: Auditory (cochlea) and visual (retina)
  cognitive processing applications,'' in \emph{2012 International Electron
  Devices Meeting}, Dec 2012, pp. 10.3.1--10.3.4, doi:
  10.1109/IEDM.2012.6479017.

\bibitem{b8}
S.~Yu, Y.~Wu, R.~Jeyasingh, D.~Kuzum, and H.~S.~P. Wong, ``An electronic
  synapse device based on metal oxide resistive switching memory for
  neuromorphic computation,'' \emph{IEEE Transactions on Electron Devices},
  vol.~58, no.~8, pp. 2729--2737, Aug 2011, doi: 10.1109/TED.2011.2147791.

\bibitem{b5}
M.~Suri, O.~Bichler, D.~Querlioz, O.~Cueto, L.~Perniola, V.~Sousa,
  D.~Vuillaume, C.~Gamrat, and B.~DeSalvo, ``Phase change memory as synapse for
  ultra-dense neuromorphic systems: Application to complex visual pattern
  extraction,'' in \emph{2011 International Electron Devices Meeting}, Dec
  2011, pp. 4.4.1--4.4.4, doi: 10.1109/IEDM.2011.6131488.

\bibitem{b6}
G.~Srinivasan, A.~Sengupta, and K.~Roy, ``Magnetic tunnel junction based
  long-term short-term stochastic synapse for a spiking neural network with
  on-chip stdp learning,'' \emph{Scientific reports}, vol.~6, p. 29545, July
  2016, doi: 10.1038/srep29545.

\bibitem{sengupta2016hybrid}
A.~Sengupta, A.~Banerjee, and K.~Roy, ``Hybrid spintronic-cmos spiking neural
  network with on-chip learning: Devices, circuits, and systems,''
  \emph{Physical Review Applied}, vol.~6, no.~6, p. 064003, 2016.

\bibitem{b7}
E.~Covi, S.~Brivio, A.~Serb, T.~Prodromakis, M.~Fanciulli, and S.~Spiga,
  ``Analog memristive synapse in spiking networks implementing unsupervised
  learning,'' \emph{Frontiers in Neuroscience}, vol.~10, p. 482, Oct 2016, doi:
  10.3389/fnins.2016.00482.

\bibitem{b9}
A.~Fert, V.~Cros, and J.~Sampaio, ``Skyrmions on the track,'' \emph{Nature
  nanotechnology}, vol.~8, no.~3, pp. 152--156, Mar 2013, doi
  :10.1038/nnano.2013.29.

\bibitem{b25}
W.~Kang, Y.~Huang, X.~Zhang, Y.~Zhou, and W.~Zhao, ``Skyrmion-electronics: An
  overview and outlook,'' \emph{Proceedings of the IEEE}, vol. 104, no.~10, pp.
  2040--2061, Oct 2016, doi: 10.1109/JPROC.2016.2591578.

\bibitem{b24}
S.~Heinze, K.~Von~Bergmann, M.~Menzel, J.~Brede, A.~Kubetzka, R.~Wiesendanger,
  G.~Bihlmayer, and S.~Bl{\"u}gel, ``Spontaneous atomic-scale magnetic skyrmion
  lattice in two dimensions,'' \emph{Nature Physics}, vol.~7, no.~9, pp.
  713--718, July 2011, doi: 10.1038/nphys2045.

\bibitem{b23}
S.~M{\"u}hlbauer, B.~Binz, F.~Jonietz, C.~Pfleiderer, A.~Rosch, A.~Neubauer,
  R.~Georgii, and P.~B{\"o}ni, ``Skyrmion lattice in a chiral magnet,''
  \emph{Science}, vol. 323, no. 5916, pp. 915--919, 2009.

\bibitem{b18}
X.~Zhang, M.~Ezawa, and Y.~Zhou, ``Magnetic skyrmion logic gates: conversion,
  duplication and merging of skyrmions,'' \emph{Scientific reports}, vol.~5, p.
  9400, Mar 2015, doi: 10.1038/srep09400.

\bibitem{b17}
X.~Zhang, Y.~Zhou, M.~Ezawa, G.~Zhao, and W.~Zhao, ``Magnetic skyrmion
  transistor: skyrmion motion in a voltage-gated nanotrack,'' \emph{Scientific
  reports}, vol.~5, p. 11369, June 2015, doi: 10.1038/srep11369.

\bibitem{b19}
W.~Kang, C.~Zheng, Y.~Huang, X.~Zhang, Y.~Zhou, W.~Lv, and W.~Zhao,
  ``Complementary skyrmion racetrack memory with voltage manipulation,''
  \emph{IEEE Electron Device Letters}, vol.~37, no.~7, pp. 924--927, July 2016,
  doi: 10.1109/LED.2016.2574916.

\bibitem{li2017magnetic}
S.~Li, W.~Kang, Y.~Huang, X.~Zhang, Y.~Zhou, and W.~Zhao, ``Magnetic
  skyrmion-based artificial neuron device,'' \emph{Nanotechnology}, vol.~28,
  no.~31, p. 31LT01, 2017.

\bibitem{chen2018compact}
X.~Chen, K.~Wang, D.~Zhu, X.~Zhang, N.~Lei, Y.~Zhang, Y.~Zhou, and W.~Zhao, ``A
  compact skyrmionic leaky-integrate-fire spiking neuron device,''
  \emph{Nanoscale}, 2018.

\bibitem{b26}
\BIBentryALTinterwordspacing
Y.~Huang, W.~Kang, X.~Zhang, Y.~Zhou, and W.~Zhao, ``Magnetic skyrmion-based
  synaptic devices,'' \emph{Nanotechnology}, vol.~28, no.~8, p. 08LT02, Jan
  2017. [Online]. Available:
  \url{http://stacks.iop.org/0957-4484/28/i=31/a=31LT01}
\BIBentrySTDinterwordspacing

\bibitem{ambrogio2016neuromorphic}
S.~Ambrogio, S.~Balatti, V.~Milo, R.~Carboni, Z.-Q. Wang, A.~Calderoni,
  N.~Ramaswamy, and D.~Ielmini, ``Neuromorphic learning and recognition with
  one-transistor-one-resistor synapses and bistable metal oxide rram,''
  \emph{IEEE Transactions on Electron Devices}, vol.~63, no.~4, pp. 1508--1515,
  2016.

\bibitem{qiao2015reconfigurable}
N.~Qiao, H.~Mostafa, F.~Corradi, M.~Osswald, F.~Stefanini, D.~Sumislawska, and
  G.~Indiveri, ``A reconfigurable on-line learning spiking neuromorphic
  processor comprising 256 neurons and 128k synapses,'' \emph{Frontiers in
  neuroscience}, vol.~9, p. 141, 2015.

\bibitem{livi2009current}
P.~Livi and G.~Indiveri, ``A current-mode conductance-based silicon neuron for
  address-event neuromorphic systems,'' in \emph{Circuits and systems, 2009.
  ISCAS 2009. IEEE international symposium on}.\hskip 1em plus 0.5em minus
  0.4em\relax IEEE, 2009, pp. 2898--2901.

\bibitem{liu2002analog}
S.-C. Liu, \emph{Analog VLSI: circuits and principles}.\hskip 1em plus 0.5em
  minus 0.4em\relax MIT press, 2002.

\bibitem{nair2017differential}
M.~V. Nair, L.~K. Muller, and G.~Indiveri, ``A differential memristive synapse
  circuit for on-line learning in neuromorphic computing systems,'' \emph{Nano
  Futures}, vol.~1, no.~3, p. 035003, 2017.

\bibitem{mastisk}
T.~Bhattacharya, V.~Parmar, and M.~Suri, ``Mastisk: Simulation framework for
  design exploration of neuromorphic hardware,'' \emph{2018 International Joint
  Conference on Neural Networks (IJCNN)}, in press.

\bibitem{sampaio2013nucleation}
J.~Sampaio, V.~Cros, S.~Rohart, A.~Thiaville, and A.~Fert, ``Nucleation,
  stability and current-induced motion of isolated magnetic skyrmions in
  nanostructures,'' \emph{Nature nanotechnology}, vol.~8, no.~11, pp. 839--844,
  2013.

\bibitem{gould2004tunneling}
C.~Gould, C.~R{\"u}ster, T.~Jungwirth, E.~Girgis, G.~Schott, R.~Giraud,
  K.~Brunner, G.~Schmidt, and L.~Molenkamp, ``Tunneling anisotropic
  magnetoresistance: a spin-valve-like tunnel magnetoresistance using a single
  magnetic layer,'' \emph{Physical review letters}, vol.~93, no.~11, p. 117203,
  2004.

\bibitem{b22}
J.~W. Lau, R.~D. McMichael, and M.~J. Donahue, ``Implementation of
  two-dimensional polycrystalline grains in object oriented micromagnetic
  framework,'' \emph{NIST J. Res.}, vol. 114, no.~1, pp. 57--67, 2009, doi:
  10.6028/jres.114.005.

\bibitem{wiesendanger2016nanoscale}
R.~Wiesendanger, ``Nanoscale magnetic skyrmions in metallic films and
  multilayers: a new twist for spintronics,'' \emph{Nature Reviews Materials},
  vol.~1, p. 16044, 2016.

\bibitem{julliere1975tunneling}
M.~Julliere, ``Tunneling between ferromagnetic films,'' \emph{Physics letters
  A}, vol.~54, no.~3, pp. 225--226, 1975.

\bibitem{b27}
M.-H. Nguyen, D.~C. Ralph, and R.~A. Buhrman, ``Spin torque study of the spin
  hall conductivity and spin diffusion length in platinum thin films with
  varying resistivity,'' \emph{Phys. Rev. Lett.}, vol. 116, p. 126601, Mar
  2016, doi: 10.1103/PhysRevLett.116.126601.

\bibitem{sung2018effect}
C.~Sung, S.~Lim, H.~Kim, T.~Kim, K.~Moon, J.~Song, J.-J. Kim, and H.~Hwang,
  ``Effect of conductance linearity and multi-level cell characteristics of tao
  x-based synapse device on pattern recognition accuracy of neuromorphic
  system,'' \emph{Nanotechnology}, 2018.

\bibitem{abbott1993asynchronous}
L.~Abbott and C.~van Vreeswijk, ``Asynchronous states in networks of
  pulse-coupled oscillators,'' \emph{Physical Review E}, vol.~48, no.~2, p.
  1483, 1993.

\bibitem{gupta2009hebbian}
A.~Gupta and L.~N. Long, ``Hebbian learning with winner take all for spiking
  neural networks,'' in \emph{Neural Networks, 2009. IJCNN 2009. International
  Joint Conference on}.\hskip 1em plus 0.5em minus 0.4em\relax IEEE, 2009, pp.
  1054--1060.

\bibitem{b14}
D.~Heeger, ``Poisson model of spike generation,'' \emph{Handout, University of
  Standford}, vol.~5, pp. 1--13, 2000, [Online].
  Available:http://neuro.bstu.by/ai/To-dom/My$\_$research/Papers-2.1-done/LIF/Feedforward-copy-from-elsewhere/B/Poisson-spiki-train/poisson.pdf.

\bibitem{diehl2015unsupervised}
P.~U. Diehl and M.~Cook, ``Unsupervised learning of digit recognition using
  spike-timing-dependent plasticity,'' \emph{Frontiers in computational
  neuroscience}, vol.~9, p.~99, 2015.

\bibitem{jo2010nanoscale}
S.~H. Jo, T.~Chang, I.~Ebong, B.~B. Bhadviya, P.~Mazumder, and W.~Lu,
  ``Nanoscale memristor device as synapse in neuromorphic systems,'' \emph{Nano
  letters}, vol.~10, no.~4, pp. 1297--1301, 2010.

\bibitem{kuzum2011nanoelectronic}
D.~Kuzum, R.~G. Jeyasingh, B.~Lee, and H.-S.~P. Wong, ``Nanoelectronic
  programmable synapses based on phase change materials for brain-inspired
  computing,'' \emph{Nano letters}, vol.~12, no.~5, pp. 2179--2186, 2011.

\bibitem{woo2016improved}
J.~Woo, K.~Moon, J.~Song, S.~Lee, M.~Kwak, J.~Park, and H.~Hwang, ``Improved
  synaptic behavior under identical pulses using alo x/hfo 2 bilayer rram array
  for neuromorphic systems,'' \emph{IEEE Electron Device Letters}, vol.~37,
  no.~8, pp. 994--997, 2016.

\bibitem{romming2015field}
N.~Romming, A.~Kubetzka, C.~Hanneken, K.~von Bergmann, and R.~Wiesendanger,
  ``Field-dependent size and shape of single magnetic skyrmions,''
  \emph{Physical review letters}, vol. 114, no.~17, p. 177203, 2015.

\end{thebibliography}

%






\end{document}